\documentclass[reprint,amsmath,amssymb,aps,prb,groupedaddress,nofootinbib,twocolumn]{revtex4-1}
\usepackage{graphicx}
\usepackage{amsthm,amssymb,amsmath,braket,mathdots}
\usepackage{bm}
\usepackage[hidelinks,pagebackref=false,pdfnewwindow=true]{hyperref} %\hypersetup{draft}
\usepackage{epstopdf,psfrag}
\usepackage{relsize,amsbsy}
\usepackage[export]{adjustbox}

\usepackage{graphicx,xcolor,tikz}

\newcommand{\be}{\begin{equation}}
\newcommand{\ee}{\end{equation}}

\newcommand{\bit}{\begin{enumerate}}
	\newcommand{\eit}{\end{enumerate}}

\definecolor{bananayellow}{rgb}{1.0, 0.88, 0.21}
\definecolor{straw}{rgb}{0.32, 0.28, 0.1}

\begin{document}

% Title
\title{A Field Guide to Spin Liquids}

\author{Johannes Knolle}
		\affiliation{\small Blackett Laboratory, Imperial College London, London SW7 2AZ, United Kingdom}
		\author{Roderich Moessner}
		\affiliation{\small Max-Planck-Institut fur Physik komplexer Systeme, Nothnitzer Str. 38, 01187 Dresden, Germany}
		\date{\today} 

%Abstract
\begin{abstract}
Spin liquids are collective phases of quantum matter which have eluded discovery in correlated magnetic materials for over half a century. Theoretical models of these enigmatic topological phases are no longer in short supply. In experiment there also exist plenty of
promising candidate materials for their realisation. One of the central challenges for the clear diagnosis of a spin liquid has been to connect 
the two. From that perspective, this review discusses characteristic features in experiment, resulting from  the unusual properties of spin liquids. This takes us to thermodynamic, spectroscopic, transport, and other experiments on a search for traces of emergent gauge fields, spinons, Majorana Fermions and other 
fractionalised particles.
\end{abstract}

\maketitle

\tableofcontents

\section{The quest for spin liquids}
The search for spin liquids as fundamentally new states of matter is a long-running quest~\cite{Anderson,Mossner2001resonating,WenBook,Lee2008end}. Their occurence 
in insulating magnets appears to be greatly facilitated by frustrated interactions for which 
the corresponding classical spin systems display a large ground state degeneracy because the local energetics cannot be minimized in a unique way~\cite{Wannier1950antiferromagnetism,Anderson1956ordering,Villain1979insulating,Chandra1990Ising,Chubukov1992order,Moessner1998properties}. For a long time, the central and defining
concept involved was a negative one -- a (ground) state without any magnetic order -- in contrast to the prevailing phases with spontaneous symmetry breaking characterised by local order parameters. 
The rejuvenated interest in resonating valence bond (RVB) physics through high-temperature 
superconductors~\cite{Anderson1987resonating} focused attention on topological properties~\cite{Kalmeyer1987equivalence,Kivelson1987topology,Wen1989vacuum,Wen1991mean}.

Initially, theoretical ideas were centred around wave-functions, e.g. of the RVB type, but it took considerable time until microscopic Hamiltonians realising such  states at isolated points~\cite{Rokhsar1988superconductivity} or extended bona fide quantum spin liquid (QSL) phases 
were established~\cite{Mossner2001resonating}. The advent of exactly soluble model Hamiltonians~\cite{Rokhsar1988superconductivity,Kitaev2006anyons} has allowed to gain an unprecedented understanding of ground- and excited state properties of QSLs. Nowadays, the theory community
has developed a remarkable capacity to invent elaborate schemes with a plethora of different phenomenologies but arguably with little guidance from
experiment. At the same time, while the target space of interesting models has exploded, the arsenal of methods
for their detection has  not grown commensurably. Nonetheless, there has been a sustained materials physics effort
covering a huge number of magnetic compounds, and many a promising candidate systems have been unearthed, some
of which have benefitted from an intense research effort for clarifying their properties in considerable detail. 

The aim of this review is to contribute towards redressing this balance. In particular, we discuss the rich phenomenology of spin liquids in order to make connection to past and future experiments. 

Before we embark on this, we would like to motivate this programme with a few words about how it fits in with
the broader research landscape of modern condensed matter physics. The search for spin liquids nowadays forms
part of the grand challenge of understanding existence, scope and nature of physics  
beyond the standard theory of Landau and spontaneous symmetry breaking. As such, they fall in the field of 
topological condensed matter physics under the headings of long-range entangled phases and topological order. 

To this date, we have only one outstanding established class of experimental systems 
(at least in terms of materials science and beyond one dimension) exhibiting a topologically ordered quantum phase with fractionalized excitations:  this is 
the fractional quantum Hall effect. The Laughlin state and its even more elaborate brethren~\cite{Moore1991nonabelions,Read1999beyond}
have attracted much attention, most recently fuelled by the dream of realising a quantum computer topologically 
protected against decoherence~\cite{Kitaev2003fault}. This is nicely reviewed in Ref.~\cite{Nayak2008non}. 

QSLs have the added attraction of accessing the vast space of possible materials provided by the combinatorial richness 
of the periodic table, the presence of sometimes large exchange energy scales, as well as a high degree of tuneability and being amenable to experimental probes, 
e.g.\ 
through the application of magnetic fields  and the use of neutron scattering. 
The central goal for the foreseeable future therefore is an unambiguous identification of a QSL phase. For this, new experiments, including
new probes, may be needed, accompanied by a reliable theoretical analysis framework. 

Our approach here is to support this quest, but  not by providing a complete introduction to quantum spin liquids for the expert. Rather, we present a compendium of ideas to provide a broader overview, which can also act as a guide for the newcomer. Many reviews are available
with material which we do not cover here, e.g.\  comprehensive 
reviews on spin liquids \cite{Zhou2017quantum} and frustrated magnetism more generally \cite{Lacroix2011introduction}; about spin ice \cite{Bramwell2001spin}; introductory reviews to quantum spin liquids \cite{Moessner2006geometrical,Balents2010spin,Lee2008end,Mila2000quantum};  an overview over spin ice~\cite{Castelnovo2012spin} and Kitaev QSLs~\cite{Hermanns2018physics} from a fractionalisation perspective, as well as a nice review of classical emergent Coulomb gauge fields \cite{Henley2010coulomb}; and works focusing on the materials aspects of pyrochlores~\cite{Gardner2010magnetic}, herbertsmithite~\cite{Norman2016herbert} and Kitaev material candidates~\cite{Rau2016spinAnnualReview,Winter2017models}.

Physics is an experimental science. However, while discoveries are mostly driven by experiments, the resulting
insights are naturally preserved in the language of theory. The history of the search for spin liquids is therefore
naturally intertwined with what phenomena one ex- and includes under this heading. Before we move on to the core 
of this field guide, we provide a brief review of the background taxonomy. 

The remaining sections are then devoted to spin liquid phenomenology. The ultimate ambition
--  to provide a textbook, not unlike standard solid state physics textbooks on conventional phases, 
on the behaviour of such topological phases -- is a step too far for us, and we make a selection of phenomena which
we feel are particularly instructive and/or realistically attainable.

\begin{figure*}
	\includegraphics[width=1.0\linewidth]{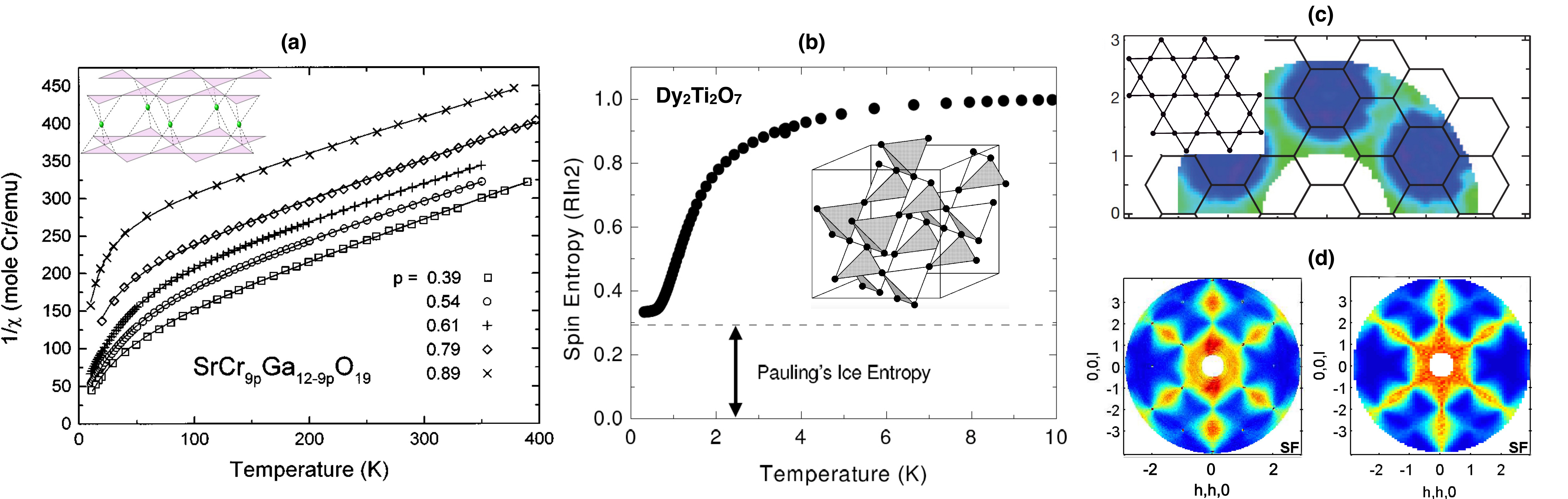}
	\caption{(Color online) (a) The Curie-Weiss temperature for various levels
	of dilution in SCGO$_x$ is much larger than the featureless regime of the susceptibility (adapted from Ref.~\cite{Schiffer1997two}). The schematic lattice structures are shown in the insets.  
	(b) The residual entropy of spin ice and the frustrated pyrochlore lattice (after Ref.~\cite{Moessner2006geometrical}. (c) Broad features
	in inelastic neutron scattering of the kagome lattice material herbertsmithite~\cite{Han2012fractionalized}. 
	(d) The emergent gauge field of 3d spin ice gives rise to pinch points in the static structure factor (left panel, neutron scattering experiment; right panel, theory; adapted from Ref.\cite{Fennell2009magnetic}).}
	\label{fig:1}
\end{figure*}

\subsection{What is a spin liquid}
What is certainly true is that the meaning of the term has shifted over the years. This is not an uncommon state of affairs, driven not only by the human tendency to adapt a definition to the requirements of the moment, but also by the fact that as the understanding of the subtleties of the phenomenon advances, refinements to the 
concepts follow.  

A constitutive concept for a spin liquid is the absence of magnetic order of a system of interacting spins at temperatures smaller than the interaction scale. This encodes the idea of a phase beyond the Landau paradigm (which covers all forms of magnetic order), as well as the intuition that a 'liquid' should be different from a 'solid'. In this sense, other forms of ordering of the spin degrees of freedom -- such as nematic orders~\cite{Shannon2006nematic,Penc2011spin} -- also a priori disqualify a system from being classified as a spin liquid. 

This negative definition, about the absence of something, continues to be the most ubiquitous and intuitive. Besides its practical limitation, to which we return below, it is nowadays considered to be somewhat too broad. 
For instance, it includes models -- interesting in their own right -- which are considered somewhat too simple. 
One is a quantum paramagnet, such as the kagome lattice Ising model in a transverse field, which is connected continuously to a high-temperature (classical) paramagnetic phase. Another is the Shastry-Sutherland model, whose two spins per structural unit cell form a dimer at low temperature, producing a simple inert state which again is straightforwardly connected to a high-temperature paramagnet but which can also display protected edge excitations of triplons at low temperatures~\cite{Mcclarty2017topological}. One calls this broader class of disordered magnets `cooperative paramagnets', to distinguish them from magnets disordered by thermal fluctuations at high temperatures, although this nomenclature is by no means universally used.

A more 'modern', positive definition involves listing conditions which a phase should meet to qualify as a spin liquid. 
This derives from advances in our understanding of what phases beyond the Landau paradigm can look like, and 
applies these to spin systems. 

Akin to the Mermin-Wagner theorem~\cite{Mermin1966absence}, which forbids spontaneous breaking of continuous symmetry (SSB) at finite temperatures in dimensions d$\leq$2, there is a rigorous result for spin systems with short range interactions. For systems with half-odd integer spin per unit cell, hence proper Mott insulators,  and without symmetry breaking, the Lieb-Schultz-Mattis theorem states~\cite{Lieb1961two} that the ground state is either unique with gapless excitations or degenerate with a gap to excitations. It establishes to some level of mathematical rigour~\cite{Oshikawa2000commensurability,Hastings2004lieb} the possibility of gapless QSLs or gapped ones with topological order. 

Perhaps the crispest is to demand that the magnet should at low temperatures
 be described by a topological field theory, such as Chern-Simons theory \cite{Zhang1989effective}.
 On the one hand, this is very restrictive, a priori ruling out
 gapless spin liquids, and spin liquids with some additional ordered degrees of freedom \cite{Brooks2014magnetic}. 
 On the other hand, this exclusion is not arbitrary--trying to 
 braid quasiparticles in the presence of gapless Goldstone modes of a ferromagnet, like for SU(2) 
 quantum Hall Skyrmions~\cite{Sondhi1993skyrmions},  does present an obstacle for envisaged quantum computation
 experiments \cite{Nayak2008non}.
 
 However, in practise to qualify as a spin liquid, it may be enough to bear in mind that 'some subset' of the degrees of freedom
 should look 'essentially topological'.  This is more or less the attitude we will take for the remainder of this
 review. For the purposes of a field guide, we will therefore be looking for {\it fractionalised excitations} and {\it emergent
 gauge fields}. The latter two are intimately linked together as standard spin flip excitations always lead to integer changes of the total spin. Therefore, excitations labeled by 'fractions' of such quantum numbers, e.g. quasiparticles carrying half-integer spin, need to be created in even numbers and once separated we can think of the emergent background gauge field as taking care of the global constraint.  The connection holds particularly in higher dimensional spin liquids, which are the focus of our field guide, but in $d=1$ fractionalisation may appear much simpler via domain wall excitations of ordered states.

\subsection{How to tell one, as a matter of principle$\ldots$}
The characterisation of a topological state of matter proceeds most easily via its global properties. Given the importance
that numerical simulations have played in advancing the field, some of these -- while appearing rather complex -- are 
comparatively straightforwardly diagnosed numerically. 

The topological order discovered by Wen and Niu posits that topological states have a degeneracy which depends
on the genus of the  surface they live on~\cite{Wen1990ground}. For instance, the Laughlin state at filling fraction $\nu=1/3$ is non-degenerate
on a sphere, and threefold degenerate on a torus. This is intimately connected with the existence of fractionalised 
quasiparticles. When a pair of Laughlin quasiparticles of charge $\pm e/3$ is created from a ground state, and one member 
of the pair moves around a non-contractible loop of the torus before annihilating the other, the
system moves from one ground state to another. Only once three such particles, and hence an electron with unit charge,
have made such a trajectory does the system return to the original ground state. Indeed, the connections between quantum
Hall physics and quantum spin liquids can become remarkably detailed, such as in transfers of wavefunctions
between the two settings, see e.g.~\cite{Schroeter2007spin}.

Observing the topological ground state degeneracy can be challenging~\cite{Yan2011spin,Jiang2012spin} but numerical methods like DMRG are well-suited for extracting an equally reliable quantifier for gapped QSLs. In analogy to long range order of conventional phases the {\it long-range entanglement} can serve as an order parameter of  topological phases~\cite{Hamma2005bipartite,Kitaev2006topological,Levin2006detecting}. The entanglement entropy of a ground state wave function can be calculated from a reduced density operator with one part of the total degrees of freedom of a bi-partitioned system traced out (with a smooth boundary of length $L$ separating the two regions). For gapped phases it follows a universal scaling form 
\begin{align}
S=c L -\gamma+...
\end{align}
The first term with a non-universal prefactor $c$ is the 'area' law common to all gapped phases but the second term $\gamma$ quantifies the long-range entanglement (it is independent of the length of the boundary of the partition). It is only nonzero in a topological phase  and is directly related to the emergent gauge structure of the QSL phase, e.g. in a $Z_2$ QSL $\gamma=\ln 2$~\cite{Levin2006detecting}. 
More detailed analyses can then yield information on the properties of fractionalised excitations and edge spectra~\cite{Cincio2013characterizing}.

\subsection{$\ldots$ and in practice}

Above, we called these diagnostics 'comparatively straightforward' because the experimental situation is
considerably less promising. The entanglement entropy -- in particular, any subleading contribution to it -- does not
correspond to any natural measurement on a many-body quantum spin system. Also, putting even a two-dimensional
magnet on a manifold of nontrivial topology sounds like a thought experiment par excellence, even more so than 
the idea of diagnosing a spin stiffness for a conventionally ordered magnet by twisting boundary conditions. 

The core aim of this review is to address precisely the question how to move forward from here. 
Lacking a silver bullet or smoking gun (or whatever
alternative martial metaphor one prefers to use) in experimental reality, one needs to make do with the probes
that actually exist, and think about how best to employ and combine them for an unambiguous identification of spin liquids.

\subsection{The role of universality}
Before we turn to this in more detail, we would like to raise an additional 'ideological' point of fundamental importance
which may  at times be somewhat under-appreciated when making contact between theory and experiment. 

The tension arises from the need in theory to devise precise definitions. One of the most successful of these is the idea
of universality, which is intimately related to the success of the developments of the concepts of symmetry
breaking and encoded in the renormalisation group half 
a century ago. Phases and phase transitions have properties which are independent of microscopic details of the 
Hamiltonian -- these properties are called universal. 

The question then is: how much of this universality is visible -- and
where/how -- in an actual experiment. The fundamentalist answer is: a priori, nothing. A case in point is the
existence of Goldstone modes accompanying the breaking of a continuous symmetry, 
in the limit of long wavelengths and low frequencies. Obviously, the limit of low frequencies will be cut off
by a finite energy  resolution -- not only due to Heisenberg's uncertainty relations -- 
of any conceivable experimental probe. On top of these, innumerable other
limitations, many of them based on nothing less than the second law of thermodynamics, will always be with us. 
We have to live with them (but can at times perhaps, see Sec.~\ref{sec:defects}, even turn them to our advantage).

In practise, this observation is not just a complaint around the edges. As we will argue below, many of the most
striking manifestations of spin liquid behaviour in fact are non-universal in the sense that they could be altered
without leaving the phase; or conversely, that proximate phases may exhibit the behaviour we are interested in 
essentially just as characteristically as the pristine version. 

As a poster child of this, we would like to adduce the fractionalised Heisenberg chain, see Fig.\ref{fig:3} (a). The agreement between theory and experiment is striking, up to considerable
detail of the structure factor at high energies, including subtle intensity variations with wavevector and frequency. However, 
none of these are universal. Close inspection of the universal part of the spectrum at low frequencies reveals the opposite~\cite{Lake2005quantum}:
due to the residual coupling between neighbouring chains, they undergo an ordering transition into {\it a different phase} with
different, 'conventional' universal behaviour of a long range ordered 3d magnet.

A fundamentalist `universalist' 
perspective therefore leaves two non-palatable alternatives: either one has the low-temperature, conventional
ordered phase; or, above the ordering transition, a phase continuously connected to a boring high-temperature paramagnet. 
Both miss the remarkable, and in our minds convincing, evidence for fractionalisation `in practise' in this compound. 
As is so often the case, while the worldview of the fundamentalist is deceptively simple, 
much of what makes real life  interesting lies in  the grey areas, the appreciation of which requires an open mind.

\section{How to start looking $\ldots$}
A time-honoured way of making a first cut at the diagnosis of spin liquidity in a candidate material is via studies of thermodynamic properties, Sect. \ref{sec:tandt}, in part because these are relatively 
easy to carry out locally in a laboratory. The first chore is to establish the absence of magnetic ordering in 
a strongly interacting (low-temperature) regime. 

A popular measure for its existence is the 
frustration parameter $f=|\Theta_{CW}|/T_f$ \cite{Obradors1988magnetic,Ramirez1994strongly},
which is defined as the ratio of two quantities. One is the Curie(-Weiss) temperature extracted from a straightforward fit
to the high-temperature susceptibility, which to first order in a high-temperature expansion is given by 
$\chi=C/(T-\Theta_{CW})$. This expression applies to an insulating magnet the size and nature of whose magnetic moments
determine the Curie constant $C$, and whose interactions determine the size of $\Theta_{CW}$. The second quantity, $T_f$, is the
location of any non-analyticity (divergence, cusp, $\ldots$) 
in $\chi$, indicating a residual ordering tendency, or spin freezing which is commonly encountered
in frustrated magnets. The regime in temperature $\Theta_{CW}\gg T>T_f$, the cooperative paramagnetic
regime, is then a natural place to start looking for a spin liquid, and it is well defined provided $f$ is sufficiently large, see Fig.\ref{fig:1} (a) for a classic example. 

If appropriate measurements are possible and available (e.g.\ thanks to the availability of a neutron source), the
absence of ordering can be confirmed by verifying that  no magnetic Bragg peaks appear when cooling down the system. 

The challenge in practice lies in the need to eliminate the presence of less obvious ordering tendencies (such as
multipolar or distortive order), which are more elusive in neutron scattering; and also not to miss
any features in the specific heat, which is nonspecific in the type of orderings it picks up, other than 
requiring the corresponding non-analytic features of the phase transition to be discernible above its smooth background temperature dependence. Also, it may be polluted by `incidental' phase transitions, such as those of the lattice which have little bearing on its magnetism.

Further valuable insights can be gleaned in  spectroscopic experiments, Sec.~\ref{sec:spec} and Fig.\ref{fig:3}. 
These can provide considerably more
detailed information than the purely macroscopic thermodynamic ones: inelastic neutron scattering
has made tremendous technological progress in the past few
years, and now routinely provides 
data in $d+1$ (wavevector+frequency) space. Other probes, such as Raman scattering, and NMR lack wavevector dependence and provide a local response averaged over the full system. 
Crucially, alternative probes each couple differently to the magnetic degrees of freedom, and therefore, they come with different selection rules for probing 
the quasiparticles (or non-quasiparticle excitations) of the material thereby providing  complementary evidence, as we discuss below. 

One central motivation for the use of finite-frequency probes lies in the fact that the ground states
of topological systems are at first sight unspectacular. 
While gapless spin liquids should generically come with algebraically decaying ground state correlations,
gapped spin liquids (just like the Laughlin charge liquid) really look featureless. 
It is the fractionalised quasiparticles which provide  a local indication of the topological physics
involved: instead of the magnons in an ordered magnet, one looks for spinons, holons, 
monopoles and the like. (An alternative to studying such finite-frequency responses lies in enlisting the help of
disorder to nucleate these excitations already in the ground state, Sec.~\ref{sec:defects}.)

For such excitations, kinematic considerations can play an important role: 
broad responses for a scattering experiment involving the creation of several particles, 
thereby weakening the restrictions imposed by energy and momentum conservation, are
taken as a prime indicator of novel spin liquid physics. In the case where the excitations of
the emergent gauge field are very heavy -- as in the case of Kitaev's QSL, where they do not
move at all -- they can effectively remove momentum conservation as a kinematic constraint, 
as their energy is barely momentum dependent~\cite{Punk2014topological}, Fig.\ref{fig:1}(c). 
In other cases, there may 
be direct evidence for the emergent gauge field, as is provided by the pinchpoints in spin ice \cite{Castelnovo2012spin},
see Fig.\ref{fig:1}(d).

\subsection{$\dots$ and where} %better in the 'field guide jargon': ... where is the habitat
The search for spin liquid compounds has been going on for a long time, and many compounds have 
yielded interesting insights and phenomena. To conclude this introduction, we mention some
materials which have shaped our own thinking -- this article tries to synthesize the resulting insights, rather
than discuss and model each system -- or indeed, all systems -- in specific detail. 

Kagome-based lattices are prime examples for geometric frustration and very prominent for quasi two-dimensional compounds. The jarosites~\cite{Wills2001conventional} 
represent a family of compounds with varying degrees of further-neighbour interactions and disorder. Beyond this,
two particularly well-studied
systems are volborthite~\cite{Hiroi2001spin}, which is now believed to have an important spatial anisotropy~\cite{Yavorski2007heisenberg,Nilsen2011pair} and 
herbertsmithite~\cite{Helton2007spin,Mendels2007quantum}. The latter is in some ways  an outstanding candidate~\cite{Norman2016herbert}. 
Next in line are triangular compounds, where a family of organic systems ('dmits') are particularly prominent~\cite{Shimizu2003spin},
as discussed in the thermodynamics section below. 

In the last few years, candidate materials~\cite{Jackeli2009,Chaloupka2010kitaev} for Kitaev spin liquids have received intense attention \cite{Rau2016spinAnnualReview,Winter2017models,Hermanns2018physics}, with a combination of mapping out the Hamiltonian and understanding the resulting consequences keeping a large community busy. These include the 'Kitaev Iridates' A$_2$ IrO$_3$ with A$=$Na,Li~\cite{singh2010antiferromagnetic,Singh2012relevance}, as well as $\alpha$-RuCl$_3$~\cite{Plumb2014rucl3,Banerjee2016proximate}, see Fig.\ref{fig:3}(b+c).

Historically very important -- as the founding material of highly frustrated magnetism -- has been SCGO \cite{Obradors1988magnetic}, a kagome-triangle-kagome
trilayer, which may alternatively be viewed as a slab of the pyrochlore lattice of corner-sharing tetrahedra. The pyrochlore
lattice in turn hosts the Ising magnet known as spin ice~\cite{Bramwell2001spin,Castelnovo2012spin}, 
most prominently or Dy/Ho$_3$Ti$_2$O$_7$,  
the only generally acknowledged fractionalised magnetic material in 
three dimensions. Many other compounds exist on this lattice, such as a more 'quantum' version with $Pr$ ions \cite{Wen2017disordered},
as well as a large class of spinel compounds, including the well-studied Cr-spinels~\cite{Lee2000local} with, like SCGO, isotropic $S=3/2$
moments. Diluting a quarter of sites of that lattice in turn yields the hyperkagome lattice, with Na$_4$Ir$_3$O$_3$ 
the most prominent exponent~\cite{Okamoto2007spin}. 

Newcomers are constantly added to this list, most recently Ca$_{10}$Cr$_7$0$_28$~\cite{Balz2016physical}, Ba$_3$NiSb$_2$O$_9$~\cite{Cheng2011high}, YbMgGaO$_4$~\cite{Paddison2017continuous}, Cu$_2$IrO$_3$~\cite{Abramchuk2017cu2iro3} and H$_3$LiIr$_2$O$_6$~\cite{Kitagawa2018spin}.

\section{Thermodynamics and transport}
\label{sec:tandt}
A key target for diagnosing spin liquids is the experimental identification of fractionalized excitations at low energies. Thermodynamic and transport measurements are complementary  in this endeavour, the former probing the necessary low energy density of states (DOS) and the latter the mobility of the excitations. 
The absence of standard Goldstone modes from conventional symmetry breaking phases, e.g. spin waves of an ordered magnet,  can be deduced from the the absence of non-analyticities in thermodynamic observables. In practice many material candidates have preemptive symmetry breaking instabilities leading to non-analyticities from sub-leading interactions. It prevents a true low temperature liquid phase for example due to weak interlayer couplings of quasi-two-dimensional materials, but as long as the frustration parameter $f$ is small enough the correlated paramagnetic regime at intermediate temperatures  is a good starting point in the search for QSL physics. 

Candidate spin liquids often exhibit a {\it spectral weight downshift} of the specific heat as part of their refusal 
to order. This can in the most extreme cases go as far as apparent violations of the third law of thermodynamics. This 
happens in spin ice, Fig.\ref{fig:1} (b), where upon cooling a 'residual entropy' is measured, which indicates that even at the lowest
temperatures, the system continues to explore an exponentially large number of states. This in particular sets
cooperative paramagnetism apart from, say, dimensionality-induced destruction of ordering. While purely one-dimensional
spin systems such as a $S=1/2$ Heisenberg antiferromagnetic chain 
do not order at any nonzero temperature, they nonetheless are close to an ordered state and often lose most of their 
entropy already upon cooling through $\Theta_{CW}$.

\begin{figure*}
        \includegraphics[width=1.0\linewidth]{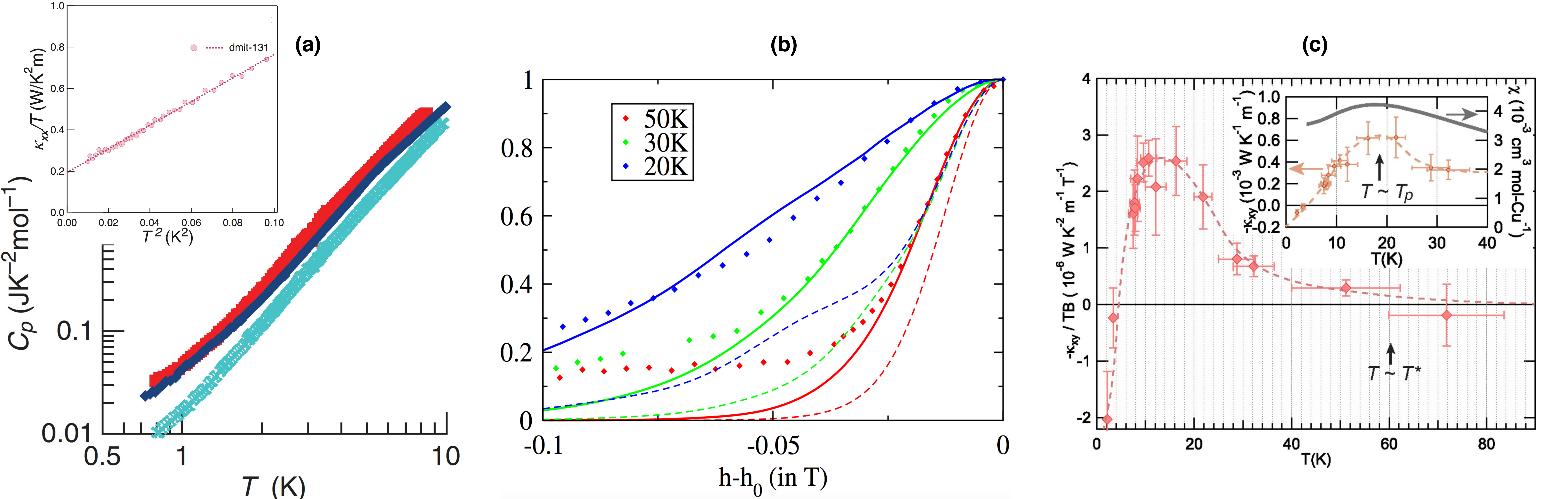}
       	\caption{(Color online) (a)  Linear specific heat~\cite{Yamashita2011gapless} and residual linear in temperature longitudinal heat conductivity~\cite{Yamashita2010highly} (inset) in triangular organic compound 'dmit'. 
	(b) Linewidth in NMR experiment of SCGO due to orphan spins and their interactions~\cite{Sen2011fractional,Limot2002susceptibility}. (c) Evidence for a thermal Hall effect in the kagome volborthite~\cite{Watanabe2016emergence}.}	
	\label{fig:2}
\end{figure*}

\subsection{Thermodynamics}
Even though characteristic correlations of a QSL are only expected at temperatures well below  $\Theta_{CW}$, the high temperature thermodynamic response of a candidate material  already contains useful information. Details of a microscopic description are obtained by a comparison of the temperature scale at which (and the form of how) the magnetic susceptibility deviates from Curie-Weiss behaviour to that calculated in a high-temperature expansion of a putative spin Hamiltonian~\cite{Oitmaa2006series,Lohmann2014tenth}.  The angular magnetic field dependence of $\chi$ with respect to crystal orientation is in principle able to detect spin-anisotropic interactions~\cite{Modic2014Realization}. The goal is to invert macroscopic measurements to microscopic descriptions which however is only attainable as long as the low energy spin Hamiltonian and the magnetic $g$-tensor are sufficiently simple.

A defining  feature of QSLs are fractionalized magnetic excitations which, after subtracting other  contributions to the heat capacity
(mainly from phonons, but at times, in particular at the lowest energies, nuclear spin), can be probed via the low temperature dependence of thermodynamic observables. In a gapless QSL information about a low energy power-law DOS, $N(\omega) \propto \omega^{\alpha}$, can be readily extracted because the specific heat is able to directly probe the  exponent $\alpha$   
\begin{align}
\frac{C_V}{T} & =  \frac{1}{T}\frac{\partial}{\partial T} \int \text{d}\omega \omega N(\omega ) n(\omega,T) \propto T^{\alpha}.
\end{align}
Hence, in conjunction with the dimensionality of the system detailed information about the low energy dispersion can be inferred such as the presence of emergent Fermi surfaces, Dirac points or nodal lines, all of which have been proposed in model QSLs~\cite{Wen2002quantum,Hermele2004stability,Obrien2016classification}. 
Of course, the procedure rests on assumptions like the presence of weakly interacting quasi-particles whose thermal distribution only depends on their individual energies, for example $n(\omega,T)$ being the Fermi-Dirac or Bose-Einstein distribution functions. 

Another caveat is that these power laws are not necessarily fixed in all circumstances. For instance.\ in a honeycomb system
with slowly varying bond disorder the resulting hopping problem is equivalent to particles in a random gauge field. Famously, 
this problem gives rise to a density of states with a power increasing {\it continuously} with disorder strength~\cite{Ludwig1994integer}. 
Such strains may quite conceivably be present in, say, organic systems with relatively low lattice rigidities.

Another intrinsic complication for a straightforward interpretation of thermodynamic data could be the presence of very different energy scales making it hard to estimate the right scaling regime. For example, the Dirac spectrum of the honeycomb Kitaev QSL with $N(\omega)\propto \omega$ would simply predict $C_V/T\propto T$ which turns out to be only observable at extremely low temperatures but instead over a large temperature window a metallic like $C_V/T\propto T^0$ appears~\cite{Nasu2014vaporization,Nasu2015thermal}. The reason is that spin flip excitations fractionalize into Majorana fermions and flux excitations.  The latter have a small gap which is only a fraction of the total magnetic energy scale. At all but the lowest temperatures the presence of thermally excited fluxes destroys the Dirac spectrum of the Majorana fermions changing the low energy DOS to roughly a constant $N(\omega)\propto \omega^0$. This particular example shows, on the one hand, the difficulties of drawing reliable conclusions from thermodynamic measurements. On the other hand, it highlights how a close comparison of microscopic calculations and experimental data could be used in principle to extract complementary information about the fractionalized excitations of a QSL.

\subsection{Longitudinal transport: thermal versus charge}
In the absence of mobile charge carriers in magnetic insulating materials thermal transport experiments can probe the mobility of elementary excitations. Ideally the heat flow along a temperature gradient contains information about the velocity, $v_{\mathbf{k} }$, and mean-free path, $l_{\text{MFP}}$, of fractionalized quasi-particles of a QSL with dispersion $E_{\mathbf{k}}$, e.g. as obtained for the longitudinal thermal conductivity $\kappa$ in a semi-classical Boltzmann calculation 
\begin{align}
\kappa & =  \frac{\partial}{\partial T} \int \text{d}^dk  \ l_{\text{MFP}}  E_{\mathbf{k}} |v_{\mathbf{k} }| n(E_{\mathbf{k}},T).
\end{align}
In a number of low dimensional QSL candidate materials such purely magnetic contributions have been observed, e.g. in triangular~\cite{Yamashita2009thermal,Yamashita2010highly}, kagome~\cite{Watanabe2016emergence} and Kitaev honeycomb~\cite{Leahy2017anomalous,Hentrich2017large} systems. 
However, in general it is very hard to get rid of the spurious phonon contribution. Moreover, in the presence of sizeable spin-phonon couplings magnetic excitations scatter of phonon contributions and vice versa which makes it hard to disentangle the two. These problems could in principle be overcome by studying directly the {\it spin current} transport which is measurable via the inverse spin Hall effect as demonstrated with insulating ordered magnets~\cite{Kajiwara2010transmission}. This has recently been proposed for QSL materials~\cite{Chatterjee2015probing} but more theoretical work and experiments are needed to show whether spin transport measurements can be turned into a new tool for studying QSLs. 

\subsection{Bulk-boundary correspondence and quantization of currents}
\label{sec:bandb}
A crowning achievement would be the observation of  {\it quantised } transport signatures directly related to topological invariants, 
to rank alongside the famously quantised Hall conductivity of the fractional quantum Hall effect. 
Again, for insulating magnets, these cannot be charge transport, but nevertheless a quantization of the thermal Hall effect~\cite{Katsura2010theory}, $\kappa_{xy}$, has been predicted in certain types of QSLs with broken time reversal symmetry~\cite{Lee2015thermal}. 
The origin of the quantization~\cite{Kitaev2006anyons} can be illustrated for a chiral QSL with $\nu$ chiral edge modes (their one dimensional dispersions labeled by momentum $q$ connecting zero energy with the gapped bulk states) and at temperatures below the bulk gap $\Delta$. The current is simply determined by the product of energy, thermal occupation (here for fermionic spinons obeying Fermi-Dirac statistics) and their velocity
\begin{eqnarray}
I_{xy}&= &\nu \int_0^{\Delta} \epsilon(q)  n(\epsilon) v(q) \frac{dq}{2\pi} \\ \nonumber
&=& \nu \int_0^{\infty} \epsilon(q) \frac{1}{1+e^{\epsilon(q)/T}} \frac{d\epsilon}{dq} \frac{dq}{2\pi} 
= \nu \frac{\pi}{24} T^2.
\end{eqnarray}
Recent experiments have observed signatures of a thermal Hall effect in disordered magnetic insulators, e.g. in kagome Volbortite~\cite{Watanabe2016emergence}, pyrochlore compound Tb$_2$Ti$_2$O$_7$~\cite{Hirschberger2015large}  and the Kitaev candidate $\alpha$-RuCl$_3$~\cite{Kasahara2017thermal}. However, the unambiguous confirmation of the quantised prefactor of the temperature dependence is missing because it is yet again difficult to
disentangle from, say,  more prosaic types of heat transport. In addition, not every spin liquid
comes with quantised transport coefficient, and at any rate, different spin liquids may not be 
distinguishable in this way alone. Nevertheless, the resulting problems of uniqueness and completeness of
a classification scheme can safely be deferred until a time that such quantised 
transport has unambiguously been detected in at least two compounds.

\subsection{Unconventional phase transitions}
Despite the identification of a distinguishing characteristic of a  QSL via its topological properties -- the long-range entanglement of its ground state -- this can be of limited practical use as even some of the most paradigmatic states -- among them classical spin ice, the $Z_2$ gauge theories or the Kitaev honeycomb model -- are only zero temperature phases and no phase transition occurs when cooling down from the simple high temperature paramagnet. Nevertheless, the behaviour of short length and time scales [and here short could mean logarithmic in system size~\cite{Castelnovo2007entanglement}] can still be governed by the fractionalized excitations of the zero temperature QSL~\cite{Nasu2016fermionic}. 

The phase transitions out of topological phases can be of considerable autonomous importance. Since the first non-symmetry breaking phase transition beyond the Landau paradigm
identified by Wegner for lattice gauge theories~\cite{Wegner1971duality,Kogut1979an}, many other interesting proposals have been made. We mention these only in passing
since pinning these down in detail is even more challenging than identifying the relevant phase in itself.

The basic attraction of such transitions
is that they reflect the exotic nature of the emergent degrees of freedom. 
For instance, the Kasteleyn transition~\cite{Kasteleyn1963dimer} 
is an asymmetric transition on account of the string-like nature of an emergent U(1) gauge field: when a string has a negative
free energy per unit length, it is suppressed by a Boltzmann factor {\it in systems size} as it needs to span the entire system. Hence, none -- not even in the form
of fluctuations --  are present until
the sign of its free energy changes, upon which there is a totally conventional continuous onset of string density. In spin ice, 
where the topological spin state is
most reliably established, (a thermally rounded version of) this transition has indeed been observed~\cite{Moessner2003theory,Fennell2007pinch,Jaubert2008three}.

Spin ice also hosts a liquid-gas transition with a critical endpoint 
of the emergent magnetic monopoles as zero-dimensional defects in a three-dimensional topological 
phase \cite{Castelnovo2008magnetic}. These form a Coulomb liquid which can then be treated with methods imported
from electrochemistry such as Debye-H\"uckel theory and its extensions \cite{Huse2003coulomb,Castelnovo2011debye,Kaiser2018emergent}. 
This is an instance of non-trivial collective behaviour of the emergent degrees of freedom, which in itself
remains a largely unexplored aspect of the field, Sec.~\ref{sec:defects}. 

Much beautiful theory has been developed regarding such unusual phase transitions, including the identification of unusual 
signatures such as anomalously large anomalous exponents. A particular case in point is the possibility of deconfined quantum 
criticality~\cite{Senthil2004deconfined,Sandvik2007evidence}, where the critical point with deconfined excitations separates two symmetry-breaking
confined phases~\cite{Mossner2001resonating}.

\section{Spectroscopy}
\label{sec:spec}
The absence of Bragg peaks in zero frequency measurements probing static correlations is an alternative indicator for ruling out conventional symmetry-breaking phases in spin liquid candidate materials. The generic situation of the correlations not only lacking a long-range ordered component, but also being numerically short-range means that, in reciprocal space, all features are broad.

Inelastic scattering experiments at nonzero frequency have the big advantage of also probing excited states beyond the asymptotic low energy regime of thermodynamic measurements. The apparent disadvantage that these are 'non-universal' is remedied by the prospect of identifying concrete spin liquids in actual experiments. 
Many experimental probes with ever increasing frequency resolution are available. Each of these has its well-developed strengths
and also well-known set of shortcomings, to discuss all of which would go beyond the scope of a simple review such as this,
so that we will emphasize the points specific to spin liquids in the following. 

First of all, just like in the thermodynamic probes, one thing to fundamentally look out for is 
an unusual temperature dependence of dynamical correlations~\cite{Broholm1990antiferromagnetic}, see Fig.\ref{fig:3} (c) for a recent example. There should again be a cooperative
paramagnetic regime in which interactions are strong but response functions change little as the temperature is lowered.

\subsection{Inelastic neutron scattering and quasi-particle kinematics}
The method of choice for measuring the basic spin correlation functions -- both static and dynamic -- is inelastic neutron scattering whose cross-section is given by
\begin{align}
&\frac{\text{d} \sigma}{\text{d} \Omega \text{d} E} \propto F(q) \left( \delta_{\alpha \beta} -\frac{q^{\alpha} q^{\beta}}{q^2}\right) \times &\\ \nonumber  &\sum_{\mathbf{r_i},\mathbf{r_j}} e^{i\mathbf{q}\cdot \left( \mathbf{r_i}-\mathbf{r_j} \right)} \int \text{d}t \text{d}t' e^{-i\omega \left(t-t' \right)} \langle S^{\alpha}_{\mathbf{r_i}}(t) S^{\beta}_{\mathbf{r_j}} (t')\rangle .&
\end{align}
From the sum rules connecting static correlations to frequency integrated dynamical ones it is apparent that the absence of static Bragg peaks in spin liquids goes along with the spectral weight being found elsewhere at nonzero frequencies.

A central role in pursuing spin liquids and their concomitant fractionalisation is played by  selection rules. Most simply put,
if scattering involves a two-body process -- e.g.\ a neutron spin flip creating a magnon -- the twin constraints of energy and momentum
conservation can lead to a sharp response in the form of a single line of energy versus wavevector transfer, $\omega(\mathbf{q})$ which
represents the magnon dispersion relation. 
This simple situation is fortuitous in that the selection rules for neutron scattering off magnons permit precisely such
a matrix element. The combined facts that neutrons are well-matched to length/energy scales {\it and} quantum numbers of magnons
underpins some of their phenomenal success in the field of magnetism. 

A broader response can therefore have a number of different origins. Firstly, there may not be such
simple scattering processes available, such as in Raman scattering, where zero wavevector transfer $q=0$ requires
the creation of multiple magnetic excitations. Secondly, there may not be a simple magnon available, but rather only 
fractionalised excitations which have to be created together, thereby rendering the scattering process a many-particle one. 
Thirdly, there may not be a quasiparticle description of the low-energy spectrum in the first place, so that no dispersion 
relation $\omega(q)$ exists even as a matter of principle.  This latter situation, especially with view to gapless spin liquids,
 is perhaps the least understood at this point in time. 

Hence, while the existence of a broad dynamic scattering response is a good indicator of a correlated paramagnetic regime, it is itself not a sufficient piece of evidence 
for fractionalized excitations of a spin liquid. Nevertheless, given the large amount of information available in a full $\omega(q)$ map, it opens up the possibility
of providing a more microscopic modelling of the {\it non-universal} features discussed above.

\begin{figure*}
        \includegraphics[width=1.0\linewidth]{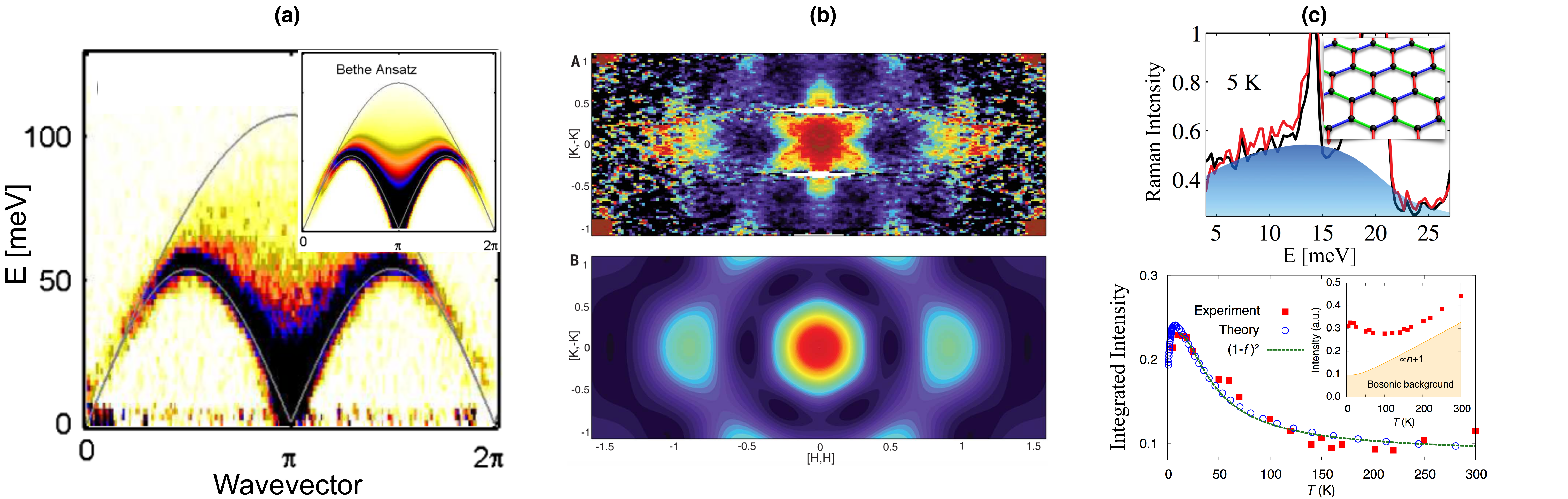}
        \caption{(Color online) Comparison of inelastic scattering experiment and theory. (a) Fractionalised spinons in the $S=1/2$ Heisenberg chain material KCuF$_3$~\cite{Lake2013multi,Mourigal2013fractional}. 
	(b) Finite frequency neutron structure factor of the proximate spin liquid in $\alpha$-RuCl$_3$~\cite{Banerjee2017neutron}.
	(c) Fermionic nature of excitations in RuCl$_3$ as evidenced in Raman scattering~\cite{Sandilands2015scattering,Nasu2016fermionic}.}	
	\label{fig:3}
\end{figure*}

For the Heisenberg chain, the kinematics of fractionalisation is most beautifully illustrated already in the three
parabolae which denote the possible energy-momentum combinations of two independent domain walls obtained
from flipping a single spin in an ordered chain. The rather nontrivial intensity map -- the kinematically allowed
processes take place with very different intensities -- follows from an actual enumeration of the matrix elements
involved, see Fig.\ref{fig:3}(a). 

In high dimensions, an analogous picture  may e.g.\ apply to a $Z_2$ spin liquid where a triplet excitation can
be decomposed into a pair of $S=1/2$ spinons, which would appear as monomers in a 
quantum dimer model~\cite{Moessner2011quantum}. For this to be straightforwardly visible, it would be necessary for the individual 
spinon to act like a coherent free particle, which is likely the case in the limit of low spinon excitation energies. However,  
it is at present unclear --  and one of the most interesting questions -- 
over what energy range such long-lived well-defined fractionalised quasiparticles will exist.

This case of symmetric fractionalisation into two particles is a particularly simple scenario. In addition, many
other emergent particles are possible, and it is a priori straightforward to come up with mean-field parton constructions
with a wide variety of different fractionalisation schemes~\cite{Wen2002quantum}. One which is not uncommon is to end up with 
a spin flip corresponding to the creation of a gauge-charged particle, and an excitation of the emergent gauge field itself, e.g. as happens in the Kitaev models~\cite{Knolle2014dynamics,Knolle2016dynamics}. This fractionalisation can be very asymmetrical: the flux may be very heavy, that is to say, have a very small bandwidth.
It can thus take up an arbitrary amount of momentum at almost constant energy, and thereby render momentum conservation 
essentially inoperative~\cite{Punk2014topological}. This then leads to  features in reciprocal space which are so broad
that it is hard to infer much about the dispersion of the light particle. Nevertheless,  key information about the energy of flux excitations and the DOS of the light fractionalized particles can in principle be inferred from the INS response~\cite{Knolle2014dynamics,Knolle2015dynamics,Knolle2016dynamics}.

\subsection{Light scattering}
Light can interact with the purely magnetic low energy degrees of freedom of Mott insulators via the virtual hopping processes, which determine the magnetic exchange constant themselves. This has been shown for example to lead to a nonzero electric polarisability. It gives rise to an AC optical conductivity in Mott insulators~\cite{Bulaevskii2008electronic} signatures of which have been analysed for a number of QSLs~\cite{Ng2007power,Potter2013mechanism,Huh2013optical,Bolens2017mechanism}. Moreover, optical absorption can directly couple via magnetic dipole excitations to spin flips and thus becomes sensitive to the zero momentum structure factor. Such experiments on a $\alpha$-RuCl$_3$ have been recently interpreted as indications of a magnetic field induced QSL~\cite{Little2017anti,Wang2017magnetic,Wellm2017signatures}. 

Alternatively, higher order photon process can induce virtual electron-hole pairs causing double spin flip excitations~\cite{Devereaux2007inelastic}. Such dynamical Raman response of kagome~\cite{Cepas2008detection,Ko2010raman} and Kitaev QSLs~\cite{Knolle2014raman,Perreault2015theory} has been analysed theoretically. Interestingly, the difference in matrix elements compared to INS permits a more direct coupling to certain types of fractionalized quasi-particles~\cite{Nasu2016fermionic} and the polarization dependence of this zero momentum probe contains additional information~\cite{Perreault2016resonant}. Hence, Raman measurements on pyrochlores~\cite{Maczka2008temperature}, herbertsmithite~\cite{Wulferding2010interplay} and two-~\cite{Sandilands2015scattering} and three-dimensional~\cite{Glamazda2016raman} Kitaev candiate materials have been interpreted in terms of the spinon DOS of QSLs but it is difficult to separate the inevitable phonon contribution to Raman scattering -- we return to this point below, Sec.~\ref{sec:stat}.
 Finally, with a further increase in energy resolution resonant inelastic X-ray scattering (RIXS) will be a promising new tool for probing QSL excitations including their momentum dependence~\cite{Halasz2016resonant,Natori2017dynamics}.

\subsection{Local probes}
Nuclear magnetic resonance (NMR) experiments probe the local magnetic fields of the spin degrees of freedom in insulators via the hyperfine interaction with nuclear levels. They are a powerful tool in the study of spin liquids~\cite{Carretta2011nmr}. For example, an NMR frequency, which remains sharp and does not split when cooling to low temperature, rules out the presence of static magnetism with inequivalent magnetic sites or a static disordered state. Even more information is obtained from the relaxation time ${1}/{T_1T}$ directly sensitive to the local magnetic susceptibility (in the zero frequency limit) related to the magnetic DOS. In a gapped QSL an Arrhenius type behaviour is expected but gapless QSLs would again lead to characteristic power law behaviour as a function of temperature similar to the specific heat but without the parasitic phonon contributions.  

An alternative probe of local magnetic fields is the relaxation of spin-polarized muons deposited in a candidate material. These $\mu$SR experiments are able to reliably distinguish between the presence of static moments due to conventional LRO, which leads to long lived oscillations of the polarization, or dynamical moments of spin liquids, which lead to a quick decay without oscillations~\cite{Yaouanc2011muon}. The main advantage of $\mu$SR is its high sensitivity but a straightforward interpretation of the data can be complicated because the positively charged muon interacts with the lattice altering the local magnetic environment. 

Unfortunately, experiments with both local {\it and} controlled spatial resolution are missing for magnetic insulators. For weakly correlated electronic materials thanks to scanning tunnelling microscopy (STM) and angle resolved photo emission spectroscopy (ARPES) a hallmark signature of topological systems -- the bulk-boundary correspondence -- could be confirmed shortly after its prediction, e.g. of the surface Dirac cone of three dimensional topological insulators~\cite{Hsieh2008topological} or of Majorana zero energy modes in superconducting wires~\cite{Mourik2012Mourik}. Of course, for a direct confirmation of topological surface states in spin liquid candidates {\it without} charged quasi-particles similar measurement tools are highly desirable. In that context it will promising to explore new directions for example spin noise spectroscopy, scanning SQUID magnetometry~\cite{Vasyukov2013scanning}, Raman microscopes~\cite{Perreault2016majorana} or inelastic scanning tunnelling microscopy~\cite{Fransson2010theory} on spin liquid candidate thin films all with {\it spatial} resolution.

\subsection{Bound states of fractionalized quasiparticles}
\label{sec:bound}
Contrary to the intuition developed from the discussion of INS experiments, 
fractionalised quasiparticles need not have a broad, fully continuous spectrum. Instead, they may form bound
or localised states. Indeed, their quantum numbers may look a lot like that of the unfractionalised spin flip; we remind the reader
that the statement of deconfinement of fractionalised particles refers to the energy cost of their separation being bounded; this
does not preclude the possibility of discrete composite states with a finite binding energy. Again, in $d=1$ 
there is a celebrated and experimentally established instance of this~\cite{Coldea2010quantum}, 
when at the magnetic field tuned critical point the exchange field leads to a discrete part of the spectrum from bound pairs of domain wall excitations.
In higher dimension,  analogous phenomena have been theoretically proposed. 
In spin ice, like in the $d=1$ case, application of a field can 
lead to bound states of monopoles with a characteristic spectrum~\cite{Wan2012quantum}.

More exotically, the possibility of the gauge field degree of freedom being involved in a ground state is present in non-Abelian spin liquids. This is analogues to the case of a $p_x+i p_y$ superconductor~\cite{Ivanov2001non} where vortices host Majorana fermion bound states, e.g. the two Majoranas of a  pair of vortices lead to a fermionic bound state whose energy goes exponentially to zero with vortex separation. Of course, the long-term goal is the controlled manipulation of the degenerate ground state manifold for braiding in the context of topological quantum computation~\cite{Nayak2008non}. Slightly less ambitious would be the observation of such a flux-Majorana bound state which has been shown to lead, for example, to a sharp contribution in the spin structure factor~\cite{Knolle2015dynamics,Knolle2016dynamics}  in the non-Abelian phase of the Kitaev honeycomb QSL in which a spin flip introduces a pair of nearest neighbour fluxes binding a pair of Majoranas below the gapped continuum response. Alternatively, already in the Abelian but {\it anisotropic} Kitaev phases~\cite{Smith2015neutron,Smith2016majorana}, the leading response can be a sharp delta-function corresponding to the addition of a pair of emergent gauge fluxes.

\subsection{Statistics}
\label{sec:stat}
The quantum statistics of quasiparticles is a very fundamental property -- it affects the many-body density of states 
even for non-interacting
particles. This is already evidenced by the (conventional) Fermi sphere and its suppressed heat capacity, and thus
in principle accessible in thermodynamic measurements already, Sec.~\ref{sec:tandt}. The temperature dependence of 
dynamical scattering experiments contains information about the {\it thermal distribution functions} which are qualitatively 
different for quasiparticles with different quantum statistics, e.g. Fermi-Dirac versus Bose-Einstein. For example, in the context of 
Raman experiments on the Kitaev candidate material $\alpha$-RuCl$_3$ a close comparison between the T-dependence of the high energy Raman response and experimental data arguably points to the presence of spin fractionalisation in terms fermionic excitations~\cite{Nasu2016fermionic}. 

In general, fractionalised quasiparticles in topological phases can have unusual exchange statistics of Anyons due to the relative phases picked up when emergent particles of different type are interchanged. Such braiding
operations are particularly of interest given their much-appreciated potential in procuring a framework for
fault-tolerant topological quantum computing \cite{Kitaev2003fault}.
Directly probing the exchange statistics of emergent particles in a quantum spin liquid is a tall order, which
at present -- given the difficulties in doing the same even 
in the much more controlled setting of quantum interferometers in the quantum Hall effect -- seems not too close at hand. 

However, there are nonetheless qualitative feature to look out for. For instance, a scattering process which creates
a set of fractionalised particles via a local interaction is sensitive to the statistics of the particles generated together,
as their relative wavefunction influences the  matrix elements for the process in question: for a pointlike interaction,
the creation of two Fermions, say, is inhibited by the vanishing of their wavefunction as the pair moves close together. 
Hence, under rather general conditions for gapped QSLs the onset of the INS response is dominated by the long range statistical interaction between Anyonic quasiparticles leading to a universal power law dependence as a function of frequency~\cite{Morampudi2017statistics}. Similarly, a promising directions will be ambitious experiments for noise spectroscopy directly probing statistics or measuring emergent quantum numbers of fractionalized quasiparticles.

\section{Disorder and defect physics}
\label{sec:defects}
The presence of disorder -- defects, vacancies, impurities and the like -- is an unavoidable fact of life in 
condensed matter systems. Indeed, in many candidate spin liquids, understanding the role of disorder is a 
crucial step towards the identification of the physics involved, see 
e.g.~\cite{Helton2007spin,Mendels2007quantum,Wen2017disordered,Savary2017disorder}.

Besides being a nuisance, however, disorder can also be used as a probe. The basic idea is that disorder can 
make fractionalisation physics visible in the ground state that would otherwise require probing 
excitations. As a simple illustration, consider the following picture. A vacancy in a two-dimensional
system can be thought
of as inserting a microscopically tiny hole into the plane -- in a sense, it changes the topology of the plane
into that of an annulus. This hole can then have effective degrees of freedom. One instance could occur 
in a non-Abelian spin liquid, where (well-separated) vacancies can host Majorana zero modes at zero energy, 
as described in Sec.\ref{sec:stat}. 
Also, not unlike impurities in a semiconductor, a vacancy can 
host a localised fractionalised excitation -- such as a 
magnetic monopole -- which would otherwise require an activation energy in the bulk~\cite{Sen2015topological}. 

In this sense, disorder physics is closely related to our discussion of dynamical probes of fractionalised degrees 
of freedom, upon replacing  dynamical probes by probes of the disorder sites. Local 
probes can then be used to resolve the signal coming from the defects. While in a bulk probe, a signal
from a small density of defects, unless it is singularly large, is easily swamped by the bulk signal, a local probe
like NMR can detect the defect response at a frequency separated from  that of the bulk. 
Again, in one dimension, edge defects have provided a  beautiful picture of the physics of the gapped
Haldane chain~\cite{Hagiwara1990observation}.
This kind of study has been carried out in some
detail for gauge-charged vacancy degrees of freedom~\cite{Sen2011fractional} following
detailed NMR measurements on SCGO~\cite{Limot2002susceptibility}, see Fig.\ref{fig:2} (b). These orphan spins~\cite{Schiffer1997two}
establish an oscillating spin texture decaying like a Coulomb law; modelling this has led to the conclusion that
the level of disorder in SCGO is likely higher than that determined from the controllable non-stochiometry only. 
The  spin liquid response to disorder can thus be used to inverse-infer properties of the composition of the material itself. 

The response to disorder can also be very intuitive. For a spinon Fermi surface, one can develop an analogy to
the response of a Fermi liquid to disorder: an impurity can induce Friedel oscillations, leading to a disturbance
surrounding the impurity modulated at the Fermi wavevector, which in turn can give rise to RKKY-type interactions
between impurities~\cite{Mross2011charge}.

As mentioned above, collective defect physics is a huge field in its own right, so far little
studied. A systematic understanding of the many-body state of a finite density of defects embedded in a topological spin liquid seems like a particularly promising direction 
for the  discovery of surprising new phenomena~\cite{Bhatt1982scaling,Damle2000dynamics,Willans2011site,Savary2017disorder,Kimchi2017valence}.
 This topic is beyond the scope of the present article, and arguably merits a stand-alone treatment.

\section{Discussion and future directions}\label{sec:discussion}\label{sec:future}
After several decades of searching for quantum spin liquids in magnetic materials we are still awaiting the unambiguous sighting of this elusive state of matter. Encouragingly, recent years have seen a flurry of discoveries of new candidate materials and novel indicative signatures of liquidity, which raise the hopes that this long search will come to a successful conclusion in the not too distant future. One of the great attractions of finding a topological phase in a magnetic systems is their high tunability. For instance, magnetic fields of a few
tens of teslas in strength are becoming increasingly routinely available. They potentially push 'proximate' spin liquid candidates~\cite{Banerjee2016proximate} in the desired direction~\cite{Hentrich2018unusual} and the impressive advances in the energy-wavevector resolution of neutron experiments, for example, enable a visualisation of the field induced melting of conventional long range magnetism~\cite{Banerjee2018excitations}. A magnetic field can not only add a Zeeman term to the Hamiltonian, but also acts in the presence of spin-orbit interactions as a versatile probe, even mimicking an effectively staggered field on different sublattices which
can change the effective dimensionality of the emergent gauge field~\cite{Castelnovo2012spin}. 

In the absence of uniquely and individually compelling features of spin liquidity in a particular material,
an investigation of how a particular feature  (gap size, mode dispersion, continuum bandwidth, bound state energy)
changes as a function of tuning parameter (external fields, pressure, composition, strain) will be a crucial ingredient for assessing the validity of a particular interpretation of experimental data in terms of a spin liquid  under the merciless action of 
Ockham's razor. In that context, a joint effort of both theory and experiment continues to be called for pinning down the rich phenomenology of spin liquids. 

Besides such more detailed model-based input, methodological progress is also on the horizon. High on the wishlist are
improved local probes, especially with a resolution approaching the lattice scale, which is currently elusive for, say,
SQUID-based devices measuring local field distributions. These could then be used to probe boundary modes or
impurity susceptibilities even in insulating magnets, in the hope of emulating their 
 huge success in electronic systems with charge degrees of freedom via STM and ARPES. Another exciting development for probing fractionalized excitations 
 lies in the realm of non-equilibrium techniques, e.g. pump-probe measurements~\cite{Dean2016ultrafast,Alpichshev2015confinement} or spin echo and noise spectroscopy. The non-equilibrium physics of topological phases is a nascent field that will surely hold numerous surprises for the patient explorer.  
 
Our field guide is also subject to continual extension on account of the search for new materials. These may arise in the form of 
 bulk materials, metal organic frameworks~\cite{Yamada2017designing}, metallic Kondo systems~\cite{Nakatsuji2006metallic}, or in `artificial' settings such as nanostructured/thin film samples. In addition,
 there remains the promise~\cite{Duan2003controlling,Manmana2013topological} of one day realising new phases of spin systems in analogue cold atomic
 quantum simulators. 

In conclusion of this little field guide, it is worth recalling that in physics, 
the most interesting phenomena are often the ones which are not anticipated at all, so that the most basic suggestion
remains to produce and analyse experimental data with an open mind.

\section*{Acknowledgements}
We are grateful to John Chalker, Dima Kovrizhin and Shivaji Sondhi, with who our respective journeys into spin liquid territory commenced; 
and to all the other collaborators and discussion partners since, too numerous to mention here, who have helped us shape the view of the
field summarised above.

\bibliographystyle{unsrt}
\bibliography{references}

\begin{thebibliography}{100}

\bibitem{Anderson}
P.W. Anderson.
\newblock Resonating valence bonds - new kind of insulator.
\newblock {\em {Materials Research Bulletin}}, {8}({2}):{153--160}, {1973}.

\bibitem{Mossner2001resonating}
R.~Moessner and S.~L. Sondhi.
\newblock Resonating valence bond phase in the triangular lattice quantum dimer
  model.
\newblock {\em Phys. Rev. Lett.}, 86:1881--1884, Feb 2001.

\bibitem{WenBook}
Xiao-Gang Wen.
\newblock {\em Quantum Field Theory of Many-Body Systems}.
\newblock Oxford University Press, 2004.

\bibitem{Lee2008end}
Patrick~A. Lee.
\newblock An end to the drought of quantum spin liquids.
\newblock {\em Science}, 321(5894):1306--1307, 2008.

\bibitem{Wannier1950antiferromagnetism}
G.~H. Wannier.
\newblock Antiferromagnetism. the triangular ising net.
\newblock {\em Phys. Rev.}, 79:357--364, Jul 1950.

\bibitem{Anderson1956ordering}
P.~W. Anderson.
\newblock Ordering and antiferromagnetism in ferrites.
\newblock {\em Phys. Rev.}, 102:1008--1013, May 1956.

\bibitem{Villain1979insulating}
Jacques Villain.
\newblock Insulating spin glasses.
\newblock {\em Zeitschrift f{\"u}r Physik B Condensed Matter}, 33(1):31--42,
  Mar 1979.

\bibitem{Chandra1990Ising}
P.~Chandra, P.~Coleman, and A.~I. Larkin.
\newblock Ising transition in frustrated heisenberg models.
\newblock {\em Phys. Rev. Lett.}, 64:88--91, Jan 1990.

\bibitem{Chubukov1992order}
Andrey Chubukov.
\newblock Order from disorder in a kagom\'e antiferromagnet.
\newblock {\em Phys. Rev. Lett.}, 69:832--835, Aug 1992.

\bibitem{Moessner1998properties}
R.~Moessner and J.~T. Chalker.
\newblock Properties of a classical spin liquid: The heisenberg pyrochlore
  antiferromagnet.
\newblock {\em Phys. Rev. Lett.}, 80:2929--2932, Mar 1998.

\bibitem{Anderson1987resonating}
P.~W. Anderson.
\newblock The resonating valence bond state in la$_2$cuo$_4$ and
  superconductivity.
\newblock {\em Science}, 235(4793):1196--1198, 1987.

\bibitem{Kalmeyer1987equivalence}
V.~Kalmeyer and R.~B. Laughlin.
\newblock Equivalence of the resonating-valence-bond and fractional quantum
  hall states.
\newblock {\em Phys. Rev. Lett.}, 59:2095--2098, Nov 1987.

\bibitem{Kivelson1987topology}
Steven~A. Kivelson, Daniel~S. Rokhsar, and James~P. Sethna.
\newblock Topology of the resonating valence-bond state: Solitons and
  high-${T}_{c}$ superconductivity.
\newblock {\em Phys. Rev. B}, 35:8865--8868, Jun 1987.

\bibitem{Wen1989vacuum}
X.~G. Wen.
\newblock Vacuum degeneracy of chiral spin states in compactified space.
\newblock {\em Phys. Rev. B}, 40:7387--7390, Oct 1989.

\bibitem{Wen1991mean}
X.~G. Wen.
\newblock Mean-field theory of spin-liquid states with finite energy gap and
  topological orders.
\newblock {\em Phys. Rev. B}, 44:2664--2672, Aug 1991.

\bibitem{Rokhsar1988superconductivity}
Daniel~S. Rokhsar and Steven~A. Kivelson.
\newblock Superconductivity and the quantum hard-core dimer gas.
\newblock {\em Phys. Rev. Lett.}, 61:2376--2379, Nov 1988.

\bibitem{Kitaev2006anyons}
Alexei Kitaev.
\newblock Anyons in an exactly solved model and beyond.
\newblock {\em Annals of Physics}, 321(1):2 -- 111, 2006.

\bibitem{Moore1991nonabelions}
Gregory Moore and Nicholas Read.
\newblock Nonabelions in the fractional quantum hall effect.
\newblock {\em Nuclear Physics B}, 360(2):362 -- 396, 1991.

\bibitem{Read1999beyond}
N.~Read and E.~Rezayi.
\newblock Beyond paired quantum hall states: Parafermions and incompressible
  states in the first excited landau level.
\newblock {\em Phys. Rev. B}, 59:8084--8092, Mar 1999.

\bibitem{Kitaev2003fault}
A~Yu Kitaev.
\newblock Fault-tolerant quantum computation by anyons.
\newblock {\em Annals of Physics}, 303(1):2--30, 2003.

\bibitem{Nayak2008non}
Chetan Nayak, Steven~H. Simon, Ady Stern, Michael Freedman, and Sankar
  Das~Sarma.
\newblock Non-abelian anyons and topological quantum computation.
\newblock {\em Rev. Mod. Phys.}, 80:1083--1159, Sep 2008.

\bibitem{Zhou2017quantum}
Yi~Zhou, Kazushi Kanoda, and Tai-Kai Ng.
\newblock Quantum spin liquid states.
\newblock {\em Rev. Mod. Phys.}, 89:025003, Apr 2017.

\bibitem{Lacroix2011introduction}
Claudine Lacroix, Philippe Mendels, and Fr\'{e}d\'{e}ric Mila, editors.
\newblock {\em Introduction to Frustrated Magnetism: Materials, Experiments,
  Theory (Springer Series in Solid-State Sciences)}.
\newblock Springer, 2011 edition, January 2011.

\bibitem{Bramwell2001spin}
Steven~T. Bramwell and Michel J.~P. Gingras.
\newblock Spin ice state in frustrated magnetic pyrochlore materials.
\newblock {\em Science}, 294(5546):1495--1501, 2001.

\bibitem{Moessner2006geometrical}
Roderich Moessner and Arthur~P Ramirez.
\newblock Geometrical frustration.
\newblock {\em Phys. Today}, 59(2):24, 2006.

\bibitem{Balents2010spin}
Leon Balents.
\newblock Spin liquids in frustrated magnets.
\newblock {\em Nature}, 464(7286):199--208, 2010.

\bibitem{Mila2000quantum}
Fr{\'e}d{\'e}ric Mila.
\newblock Quantum spin liquids.
\newblock {\em European Journal of Physics}, 21(6):499, 2000.

\bibitem{Castelnovo2012spin}
C.~Castelnovo, R.~Moessner, and S.L. Sondhi.
\newblock Spin ice, fractionalization, and topological order.
\newblock {\em Annual Review of Condensed Matter Physics}, 3(1):35--55, 2012.

\bibitem{Hermanns2018physics}
M.~Hermanns, I.~Kimchi, and J.~Knolle.
\newblock Physics of the kitaev model: Fractionalization, dynamic correlations,
  and material connections.
\newblock {\em Annual Review of Condensed Matter Physics}, 9(1):17--33, 2018.

\bibitem{Henley2010coulomb}
Christopher~L. Henley.
\newblock The coulomb phase in frustrated systems.
\newblock {\em Annual Review of Condensed Matter Physics}, 1(1):179--210, 2010.

\bibitem{Gardner2010magnetic}
Jason~S. Gardner, Michel J.~P. Gingras, and John~E. Greedan.
\newblock Magnetic pyrochlore oxides.
\newblock {\em Rev. Mod. Phys.}, 82:53--107, Jan 2010.

\bibitem{Norman2016herbert}
M.~R. Norman.
\newblock Herbertsmithite and the search for the quantum spin liquid.
\newblock {\em Rev. Mod. Phys.}, 88:041002, Dec 2016.

\bibitem{Rau2016spinAnnualReview}
Jeffrey~G. Rau, Eric Kin-Ho Lee, and Hae-Young Kee.
\newblock Spin-orbit physics giving rise to novel phases in correlated systems:
  Iridates and related materials.
\newblock {\em Annual Review of Condensed Matter Physics}, 7(1):195--221, 2016.

\bibitem{Winter2017models}
Stephen~M Winter, Alexander~A Tsirlin, Maria Daghofer, Jeroen van~den Brink,
  Yogesh Singh, Philipp Gegenwart, and Roser Valenti.
\newblock Models and materials for generalized kitaev magnetism.
\newblock {\em Journal of Physics: Condensed Matter}, 29(49):493002, 2017.

\bibitem{Schiffer1997two}
P.~Schiffer and I.~Daruka.
\newblock Two-population model for anomalous low-temperature magnetism in
  geometrically frustrated magnets.
\newblock {\em Phys. Rev. B}, 56:13712--13715, Dec 1997.

\bibitem{Han2012fractionalized}
Tian-Heng Han, Joel~S Helton, Shaoyan Chu, Daniel~G Nocera, Jose~A
  Rodriguez-Rivera, Collin Broholm, and Young~S Lee.
\newblock Fractionalized excitations in the spin-liquid state of a
  kagome-lattice antiferromagnet.
\newblock {\em Nature}, 492(7429):406, 2012.

\bibitem{Fennell2009magnetic}
T.~Fennell, P.~P. Deen, A.~R. Wildes, K.~Schmalzl, D.~Prabhakaran, A.~T.
  Boothroyd, R.~J. Aldus, D.~F. McMorrow, and S.~T. Bramwell.
\newblock Magnetic coulomb phase in the spin ice ho2ti2o7.
\newblock {\em Science}, 326(5951):415--417, 2009.

\bibitem{Shannon2006nematic}
Nic Shannon, Tsutomu Momoi, and Philippe Sindzingre.
\newblock Nematic order in square lattice frustrated ferromagnets.
\newblock {\em Phys. Rev. Lett.}, 96:027213, Jan 2006.

\bibitem{Penc2011spin}
Karlo Penc and Andreas~M L{\"a}uchli.
\newblock Spin nematic phases in quantum spin systems.
\newblock In {\em Introduction to Frustrated Magnetism}, pages 331--362.
  Springer, 2011.

\bibitem{Mcclarty2017topological}
PA~McClarty, F~Kr{\"u}ger, Tatiana Guidi, SF~Parker, Keith Refson, AW~Parker,
  Dharmalingam Prabhakaran, and Radu Coldea.
\newblock Topological triplon modes and bound states in a shastry--sutherland
  magnet.
\newblock {\em Nature Physics}, 13(8):736, 2017.

\bibitem{Mermin1966absence}
N.~D. Mermin and H.~Wagner.
\newblock Absence of ferromagnetism or antiferromagnetism in one- or
  two-dimensional isotropic heisenberg models.
\newblock {\em Phys. Rev. Lett.}, 17:1133--1136, Nov 1966.

\bibitem{Lieb1961two}
Elliott Lieb, Theodore Schultz, and Daniel Mattis.
\newblock Two soluble models of an antiferromagnetic chain.
\newblock {\em Annals of Physics}, 16(3):407--466, 1961.

\bibitem{Oshikawa2000commensurability}
Masaki Oshikawa.
\newblock Commensurability, excitation gap, and topology in quantum
  many-particle systems on a periodic lattice.
\newblock {\em Phys. Rev. Lett.}, 84:1535--1538, Feb 2000.

\bibitem{Hastings2004lieb}
M.~B. Hastings.
\newblock Lieb-schultz-mattis in higher dimensions.
\newblock {\em Phys. Rev. B}, 69:104431, Mar 2004.

\bibitem{Zhang1989effective}
S.~C. Zhang, T.~H. Hansson, and S.~Kivelson.
\newblock Effective-field-theory model for the fractional quantum hall effect.
\newblock {\em Phys. Rev. Lett.}, 62:82--85, Jan 1989.

\bibitem{Brooks2014magnetic}
M.~E. Brooks-Bartlett, S.~T. Banks, L.~D.~C. Jaubert, A.~Harman-Clarke, and
  P.~C.~W. Holdsworth.
\newblock Magnetic-moment fragmentation and monopole crystallization.
\newblock {\em Phys. Rev. X}, 4:011007, Jan 2014.

\bibitem{Sondhi1993skyrmions}
S.~L. Sondhi, A.~Karlhede, S.~A. Kivelson, and E.~H. Rezayi.
\newblock Skyrmions and the crossover from the integer to fractional quantum
  hall effect at small zeeman energies.
\newblock {\em Phys. Rev. B}, 47:16419--16426, Jun 1993.

\bibitem{Wen1990ground}
X.~G. Wen and Q.~Niu.
\newblock Ground-state degeneracy of the fractional quantum hall states in the
  presence of a random potential and on high-genus riemann surfaces.
\newblock {\em Phys. Rev. B}, 41:9377--9396, May 1990.

\bibitem{Schroeter2007spin}
Darrell~F. Schroeter, Eliot Kapit, Ronny Thomale, and Martin Greiter.
\newblock Spin hamiltonian for which the chiral spin liquid is the exact ground
  state.
\newblock {\em Phys. Rev. Lett.}, 99:097202, Aug 2007.

\bibitem{Yan2011spin}
Simeng Yan, David~A. Huse, and Steven~R. White.
\newblock Spin-liquid ground state of the s = 1/2 kagome heisenberg
  antiferromagnet.
\newblock {\em Science}, 332(6034):1173--1176, 2011.

\bibitem{Jiang2012spin}
Hong-Chen Jiang, Hong Yao, and Leon Balents.
\newblock Spin liquid ground state of the spin-$\frac{1}{2}$ square
  ${J}_{1}$-${J}_{2}$ heisenberg model.
\newblock {\em Phys. Rev. B}, 86:024424, Jul 2012.

\bibitem{Hamma2005bipartite}
Alioscia Hamma, Radu Ionicioiu, and Paolo Zanardi.
\newblock Bipartite entanglement and entropic boundary law in lattice spin
  systems.
\newblock {\em Phys. Rev. A}, 71:022315, Feb 2005.

\bibitem{Kitaev2006topological}
Alexei Kitaev and John Preskill.
\newblock Topological entanglement entropy.
\newblock {\em Phys. Rev. Lett.}, 96:110404, Mar 2006.

\bibitem{Levin2006detecting}
Michael Levin and Xiao-Gang Wen.
\newblock Detecting topological order in a ground state wave function.
\newblock {\em Phys. Rev. Lett.}, 96:110405, Mar 2006.

\bibitem{Cincio2013characterizing}
L.~Cincio and G.~Vidal.
\newblock Characterizing topological order by studying the ground states on an
  infinite cylinder.
\newblock {\em Phys. Rev. Lett.}, 110:067208, Feb 2013.

\bibitem{Lake2005quantum}
Bella Lake, D~Alan Tennant, Chris~D Frost, and Stephen~E Nagler.
\newblock Quantum criticality and universal scaling of a quantum
  antiferromagnet.
\newblock {\em Nature materials}, 4(4):329, 2005.

\bibitem{Obradors1988magnetic}
X.~Obradors, A.~Labarta, A.~Isalgué, J.~Tejada, J.~Rodriguez, and M.~Pernet.
\newblock Magnetic frustration and lattice dimensionality in srcr8ga4o19.
\newblock {\em Solid State Communications}, 65(3):189 -- 192, 1988.

\bibitem{Ramirez1994strongly}
A~P Ramirez.
\newblock Strongly geometrically frustrated magnets.
\newblock {\em Annual Review of Materials Science}, 24(1):453--480, 1994.

\bibitem{Punk2014topological}
Matthias Punk, Debanjan Chowdhury, and Subir Sachdev.
\newblock Topological excitations and the dynamic structure factor of spin
  liquids on the kagome lattice.
\newblock {\em Nature Physics}, 10(4):289, 2014.

\bibitem{Wills2001conventional}
A~S Wills.
\newblock Conventional and unconventional orderings in the jarosites.
\newblock {\em Canadian Journal of Physics}, 79(11-12):1501--1510, 2001.

\bibitem{Hiroi2001spin}
Zenji Hiroi, Masafumi Hanawa, Naoya Kobayashi, Minoru Nohara, Hidenori Takagi,
  Yoshitomo Kato, and Masashi Takigawa.
\newblock Spin-1/2 kagomé-like lattice in volborthite cu3v2o7(oh)2·2h2o.
\newblock {\em Journal of the Physical Society of Japan}, 70(11):3377--3384,
  2001.

\bibitem{Yavorski2007heisenberg}
T.~Yavors'kii, W.~Apel, and H.-U. Everts.
\newblock Heisenberg antiferromagnet with anisotropic exchange on the kagom\'e
  lattice: Description of the magnetic properties of volborthite.
\newblock {\em Phys. Rev. B}, 76:064430, Aug 2007.

\bibitem{Nilsen2011pair}
G.~J. Nilsen, F.~C. Coomer, M.~A. de~Vries, J.~R. Stewart, P.~P. Deen,
  A.~Harrison, and H.~M. R\o{}nnow.
\newblock Pair correlations, short-range order, and dispersive excitations in
  the quasi-kagome quantum magnet volborthite.
\newblock {\em Phys. Rev. B}, 84:172401, Nov 2011.

\bibitem{Helton2007spin}
J.~S. Helton, K.~Matan, M.~P. Shores, E.~A. Nytko, B.~M. Bartlett, Y.~Yoshida,
  Y.~Takano, A.~Suslov, Y.~Qiu, J.-H. Chung, D.~G. Nocera, and Y.~S. Lee.
\newblock Spin dynamics of the spin-$1/2$ kagome lattice antiferromagnet
  ${\mathrm{zncu}}_{3}(\mathrm{OH}{)}_{6}{\mathrm{cl}}_{2}$.
\newblock {\em Phys. Rev. Lett.}, 98:107204, Mar 2007.

\bibitem{Mendels2007quantum}
P.~Mendels, F.~Bert, M.~A. de~Vries, A.~Olariu, A.~Harrison, F.~Duc, J.~C.
  Trombe, J.~S. Lord, A.~Amato, and C.~Baines.
\newblock Quantum magnetism in the paratacamite family: Towards an ideal
  kagom\'e lattice.
\newblock {\em Phys. Rev. Lett.}, 98:077204, Feb 2007.

\bibitem{Shimizu2003spin}
Y.~Shimizu, K.~Miyagawa, K.~Kanoda, M.~Maesato, and G.~Saito.
\newblock Spin liquid state in an organic mott insulator with a triangular
  lattice.
\newblock {\em Phys. Rev. Lett.}, 91:107001, Sep 2003.

\bibitem{Jackeli2009}
G.~Jackeli and G.~Khaliullin.
\newblock Mott insulators in the strong spin-orbit coupling limit: From
  heisenberg to a quantum compass and kitaev models.
\newblock {\em Phys. Rev. Lett.}, 102:017205, Jan 2009.

\bibitem{Chaloupka2010kitaev}
Ji\ifmmode \check{r}\else~\v{r}\fi{}\'{\i} Chaloupka, George Jackeli, and
  Giniyat Khaliullin.
\newblock Kitaev-heisenberg model on a honeycomb lattice: Possible exotic
  phases in iridium oxides ${A}_{2}{\mathrm{iro}}_{3}$.
\newblock {\em Phys. Rev. Lett.}, 105:027204, Jul 2010.

\bibitem{singh2010antiferromagnetic}
Yogesh Singh and P.~Gegenwart.
\newblock Antiferromagnetic mott insulating state in single crystals of the
  honeycomb lattice material ${\text{na}}_{2}{\text{iro}}_{3}$.
\newblock {\em Phys. Rev. B}, 82:064412, Aug 2010.

\bibitem{Singh2012relevance}
Yogesh Singh, S.~Manni, J.~Reuther, T.~Berlijn, R.~Thomale, W.~Ku, S.~Trebst,
  and P.~Gegenwart.
\newblock Relevance of the heisenberg-kitaev model for the honeycomb lattice
  iridates a$_2$iro$_3$.
\newblock {\em Phys. Rev. Lett.}, 108:127203, Mar 2012.

\bibitem{Plumb2014rucl3}
K.~W. Plumb, J.~P. Clancy, L.~J. Sandilands, V.~Vijay Shankar, Y.~F. Hu, K.~S.
  Burch, Hae-Young Kee, and Young-June Kim.
\newblock {$\ensuremath{\alpha}\ensuremath{-}{\mathrm{RuCl}}_{3}$: A spin-orbit
  assisted Mott insulator on a honeycomb lattice}.
\newblock {\em Phys. Rev. B}, 90:041112, Jul 2014.

\bibitem{Banerjee2016proximate}
A.~Banerjee, C.~A. Bridges, J.-Q. Yan, A.~A. Aczel, L.~Li, M.~B. Stone, G.~E.
  Granroth, M.~D. Lumsden, Y.~Yiu, J.~Knolle, S.~Bhattacharjee, D.~L.
  Kovrizhin, R.~Moessner, D.~A. Tennant, D.~G. Mandrus, and S.~E. Nagler.
\newblock Proximate kitaev quantum spin liquid behaviour in a honeycomb magnet.
\newblock {\em Nature Materials}, 15:733--740, 2016.

\bibitem{Wen2017disordered}
J.-J. Wen, S.~M. Koohpayeh, K.~A. Ross, B.~A. Trump, T.~M. McQueen, K.~Kimura,
  S.~Nakatsuji, Y.~Qiu, D.~M. Pajerowski, J.~R.~D. Copley, and C.~L. Broholm.
\newblock Disordered route to the coulomb quantum spin liquid: Random
  transverse fields on spin ice in
  ${\mathrm{pr}}_{2}{\mathrm{zr}}_{2}{\mathrm{o}}_{7}$.
\newblock {\em Phys. Rev. Lett.}, 118:107206, Mar 2017.

\bibitem{Lee2000local}
S.-H. Lee, C.~Broholm, T.~H. Kim, W.~Ratcliff, and S-W. Cheong.
\newblock Local spin resonance and spin-peierls-like phase transition in a
  geometrically frustrated antiferromagnet.
\newblock {\em Phys. Rev. Lett.}, 84:3718--3721, Apr 2000.

\bibitem{Okamoto2007spin}
Yoshihiko Okamoto, Minoru Nohara, Hiroko Aruga-Katori, and Hidenori Takagi.
\newblock Spin-liquid state in the $s=1/2$ hyperkagome antiferromagnet
  ${\mathrm{na}}_{4}{\mathrm{ir}}_{3}{\mathrm{o}}_{8}$.
\newblock {\em Phys. Rev. Lett.}, 99:137207, Sep 2007.

\bibitem{Balz2016physical}
Christian Balz, Bella Lake, Johannes Reuther, Hubertus Luetkens, Rico
  Sch{\"o}nemann, Thomas Herrmannsd{\"o}rfer, Yogesh Singh, ATM~Nazmul Islam,
  Elisa~M Wheeler, Jose~A Rodriguez-Rivera, et~al.
\newblock Physical realization of a quantum spin liquid based on a complex
  frustration mechanism.
\newblock {\em Nature Physics}, 12(10):942, 2016.

\bibitem{Cheng2011high}
JG~Cheng, G~Li, L~Balicas, JS~Zhou, JB~Goodenough, Cenke Xu, and HD~Zhou.
\newblock High-pressure sequence of ba 3 nisb 2 o 9 structural phases: New s= 1
  quantum spin liquids based on ni 2+.
\newblock {\em Physical review letters}, 107(19):197204, 2011.

\bibitem{Paddison2017continuous}
Joseph~AM Paddison, Marcus Daum, Zhiling Dun, Georg Ehlers, Yaohua Liu,
  Matthew~B Stone, Haidong Zhou, and Martin Mourigal.
\newblock Continuous excitations of the triangular-lattice quantum spin liquid
  ybmggao4.
\newblock {\em Nature Physics}, 13(2):117--122, 2017.

\bibitem{Abramchuk2017cu2iro3}
Mykola Abramchuk, Cigdem Ozsoy-Keskinbora, Jason~W Krizan, Kenneth~R Metz,
  David~C Bell, and Fazel Tafti.
\newblock Cu2iro3: a new magnetically frustrated honeycomb iridate.
\newblock {\em Journal of the American Chemical Society}, 139(43):15371--15376,
  2017.

\bibitem{Kitagawa2018spin}
K~Kitagawa, T~Takayama, Y~Matsumoto, A~Kato, R~Takano, Y~Kishimoto, S~Bette,
  R~Dinnebier, G~Jackeli, and H~Takagi.
\newblock A spin--orbital-entangled quantum liquid on a honeycomb lattice.
\newblock {\em Nature}, 554(7692):341, 2018.

\bibitem{Yamashita2011gapless}
Satoshi Yamashita, Takashi Yamamoto, Yasuhiro Nakazawa, Masafumi Tamura, and
  Reizo Kato.
\newblock Gapless spin liquid of an organic triangular compound evidenced by
  thermodynamic measurements.
\newblock {\em Nature communications}, 2:275, 2011.

\bibitem{Yamashita2010highly}
Minoru Yamashita, Norihito Nakata, Yoshinori Senshu, Masaki Nagata, Hiroshi~M
  Yamamoto, Reizo Kato, Takasada Shibauchi, and Yuji Matsuda.
\newblock Highly mobile gapless excitations in a two-dimensional candidate
  quantum spin liquid.
\newblock {\em Science}, 328(5983):1246--1248, 2010.

\bibitem{Sen2011fractional}
Arnab Sen, Kedar Damle, and Roderich Moessner.
\newblock Fractional spin textures in the frustrated magnet
  ${\mathrm{srcr}}_{9p}{\mathrm{ga}}_{12\ensuremath{-}9p}{\mathrm{o}}_{19}$.
\newblock {\em Phys. Rev. Lett.}, 106:127203, Mar 2011.

\bibitem{Limot2002susceptibility}
L.~Limot, P.~Mendels, G.~Collin, C.~Mondelli, B.~Ouladdiaf, H.~Mutka,
  N.~Blanchard, and M.~Mekata.
\newblock Susceptibility and dilution effects of the kagom\'e bilayer
  geometrically frustrated network: A ga nmr study of
  ${\mathrm{srcr}}_{9p}{\mathrm{ga}}_{12\ensuremath{-}9p}{\mathrm{o}}_{19}$.
\newblock {\em Phys. Rev. B}, 65:144447, Apr 2002.

\bibitem{Watanabe2016emergence}
Daiki Watanabe, Kaori Sugii, Masaaki Shimozawa, Yoshitaka Suzuki, Takeshi
  Yajima, Hajime Ishikawa, Zenji Hiroi, Takasada Shibauchi, Yuji Matsuda, and
  Minoru Yamashita.
\newblock Emergence of nontrivial magnetic excitations in a spin-liquid state
  of kagom{\'e} volborthite.
\newblock {\em Proceedings of the National Academy of Sciences},
  113(31):8653--8657, 2016.

\bibitem{Oitmaa2006series}
Jaan Oitmaa, Chris Hamer, and Weihong Zheng.
\newblock {\em Series expansion methods for strongly interacting lattice
  models}.
\newblock Cambridge University Press, 2006.

\bibitem{Lohmann2014tenth}
Andre Lohmann, Heinz-J\"urgen Schmidt, and Johannes Richter.
\newblock Tenth-order high-temperature expansion for the susceptibility and the
  specific heat of spin-$s$ heisenberg models with arbitrary exchange patterns:
  Application to pyrochlore and kagome magnets.
\newblock {\em Phys. Rev. B}, 89:014415, Jan 2014.

\bibitem{Modic2014Realization}
K.~A. Modic, Tess~E. Smidt, Itamar Kimchi, Nicholas~P. Breznay, Alun Biffin,
  Sungkyun Choi, Roger~D. Johnson, Radu Coldea, Pilanda Watkins-Curry,
  Gregory~T. McCandless, Julia~Y. Chan, Felipe Gandara, Z.~Islam, Ashvin
  Vishwanath, Arkady Shekhter, Ross~D. McDonald, and James~G. Analytis.
\newblock Realization of a three-dimensional spin-anisotropic harmonic
  honeycomb iridate.
\newblock {\em Nature Communications}, 5:4203, June 2014.

\bibitem{Wen2002quantum}
Xiao-Gang Wen.
\newblock Quantum orders and symmetric spin liquids.
\newblock {\em Phys. Rev. B}, 65:165113, Apr 2002.

\bibitem{Hermele2004stability}
Michael Hermele, T.~Senthil, Matthew P.~A. Fisher, Patrick~A. Lee, Naoto
  Nagaosa, and Xiao-Gang Wen.
\newblock Stability of $u(1)$ spin liquids in two dimensions.
\newblock {\em Phys. Rev. B}, 70:214437, Dec 2004.

\bibitem{Obrien2016classification}
Kevin O'Brien, Maria Hermanns, and Simon Trebst.
\newblock {Classification of gapless $Z_{2}$ spin liquids in three-dimensional
  Kitaev models}.
\newblock {\em Phys. Rev. B}, 93:085101, Feb 2016.

\bibitem{Ludwig1994integer}
Andreas W.~W. Ludwig, Matthew P.~A. Fisher, R.~Shankar, and G.~Grinstein.
\newblock Integer quantum hall transition: An alternative approach and exact
  results.
\newblock {\em Phys. Rev. B}, 50:7526--7552, Sep 1994.

\bibitem{Nasu2014vaporization}
Joji Nasu, Masafumi Udagawa, and Yukitoshi Motome.
\newblock Vaporization of kitaev spin liquids.
\newblock {\em Phys. Rev. Lett.}, 113:197205, Nov 2014.

\bibitem{Nasu2015thermal}
Joji Nasu, Masafumi Udagawa, and Yukitoshi Motome.
\newblock Thermal fractionalization of quantum spins in a kitaev model:
  Temperature-linear specific heat and coherent transport of majorana fermions.
\newblock {\em Phys. Rev. B}, 92:115122, Sep 2015.

\bibitem{Yamashita2009thermal}
Minoru Yamashita, Norihito Nakata, Yuichi Kasahara, Takahiko Sasaki, Naoki
  Yoneyama, Norio Kobayashi, Satoshi Fujimoto, Takasada Shibauchi, and Yuji
  Matsuda.
\newblock Thermal-transport measurements in a quantum spin-liquid state of the
  frustrated triangular magnet-(bedt-ttf) 2 cu 2 (cn) 3.
\newblock {\em Nature Physics}, 5(1):44, 2009.

\bibitem{Leahy2017anomalous}
Ian~A. Leahy, Christopher~A. Pocs, Peter~E. Siegfried, David Graf, S.-H. Do,
  Kwang-Yong Choi, B.~Normand, and Minhyea Lee.
\newblock Anomalous thermal conductivity and magnetic torque response in the
  honeycomb magnet
  $\ensuremath{\alpha}\text{\ensuremath{-}}{\mathrm{rucl}}_{3}$.
\newblock {\em Phys. Rev. Lett.}, 118:187203, May 2017.

\bibitem{Hentrich2017large}
Richard Hentrich, Anja~UB Wolter, Xenophon Zotos, Wolfram Brenig, Domenic
  Nowak, Anna Isaeva, Thomas Doert, Arnab Banerjee, Paula Lampen-Kelley,
  David~G Mandrus, et~al.
\newblock Large field-induced gap of kitaev-heisenberg paramagnons in
  $\alpha$-rucl$_3$.
\newblock {\em arXiv preprint arXiv:1703.08623}, 2017.

\bibitem{Kajiwara2010transmission}
Y~Kajiwara, K~Harii, S~Takahashi, Jun-ichiro Ohe, K~Uchida, M~Mizuguchi,
  H~Umezawa, H~Kawai, K~Ando, K~Takanashi, et~al.
\newblock Transmission of electrical signals by spin-wave interconversion in a
  magnetic insulator.
\newblock {\em Nature}, 464(7286):262, 2010.

\bibitem{Chatterjee2015probing}
Shubhayu Chatterjee and Subir Sachdev.
\newblock Probing excitations in insulators via injection of spin currents.
\newblock {\em Phys. Rev. B}, 92:165113, Oct 2015.

\bibitem{Katsura2010theory}
Hosho Katsura, Naoto Nagaosa, and Patrick~A. Lee.
\newblock Theory of the thermal hall effect in quantum magnets.
\newblock {\em Phys. Rev. Lett.}, 104:066403, Feb 2010.

\bibitem{Lee2015thermal}
Hyunyong Lee, Jung~Hoon Han, and Patrick~A. Lee.
\newblock Thermal hall effect of spins in a paramagnet.
\newblock {\em Phys. Rev. B}, 91:125413, Mar 2015.

\bibitem{Hirschberger2015large}
Max Hirschberger, Jason~W. Krizan, R.~J. Cava, and N.~P. Ong.
\newblock Large thermal hall conductivity of neutral spin excitations in a
  frustrated quantum magnet.
\newblock {\em Science}, 348(6230):106--109, 2015.

\bibitem{Kasahara2017thermal}
Y~Kasahara, K~Sugii, T~Ohnishi, M~Shimozawa, M~Yamashita, N~Kurita, H~Tanaka,
  J~Nasu, Y~Motome, T~Shibauchi, et~al.
\newblock Thermal hall effect in a kitaev spin liquid: A possible signature of
  majorana chiral edge current.
\newblock {\em arXiv preprint arXiv:1709.10286}, 2017.

\bibitem{Castelnovo2007entanglement}
Claudio Castelnovo and Claudio Chamon.
\newblock Entanglement and topological entropy of the toric code at finite
  temperature.
\newblock {\em Phys. Rev. B}, 76:184442, Nov 2007.

\bibitem{Nasu2016fermionic}
Joji Nasu, Johannes Knolle, Dima~L. Kovrizhin, Yukitoshi Motome, and Roderich
  Moessner.
\newblock Fermionic response from fractionalization in an insulating
  two-dimensional magnet.
\newblock {\em Nature Physics}, 12:912--915, 2016.

\bibitem{Wegner1971duality}
Franz~J Wegner.
\newblock Duality in generalized ising models and phase transitions without
  local order parameters.
\newblock {\em Journal of Mathematical Physics}, 12(10):2259--2272, 1971.

\bibitem{Kogut1979an}
John~B. Kogut.
\newblock An introduction to lattice gauge theory and spin systems.
\newblock {\em Rev. Mod. Phys.}, 51:659--713, Oct 1979.

\bibitem{Kasteleyn1963dimer}
Pieter~W Kasteleyn.
\newblock Dimer statistics and phase transitions.
\newblock {\em Journal of Mathematical Physics}, 4(2):287--293, 1963.

\bibitem{Moessner2003theory}
R.~Moessner and S.~L. Sondhi.
\newblock Theory of the [111] magnetization plateau in spin ice.
\newblock {\em Phys. Rev. B}, 68:064411, Aug 2003.

\bibitem{Fennell2007pinch}
T~Fennell, ST~Bramwell, DF~McMorrow, P~Manuel, and AR~Wildes.
\newblock Pinch points and kasteleyn transitions in kagome ice.
\newblock {\em Nature Physics}, 3(8):566, 2007.

\bibitem{Jaubert2008three}
L.~D.~C. Jaubert, J.~T. Chalker, P.~C.~W. Holdsworth, and R.~Moessner.
\newblock Three-dimensional kasteleyn transition: Spin ice in a [100] field.
\newblock {\em Phys. Rev. Lett.}, 100:067207, Feb 2008.

\bibitem{Castelnovo2008magnetic}
Claudio Castelnovo, Roderich Moessner, and Shivaji~L Sondhi.
\newblock Magnetic monopoles in spin ice.
\newblock {\em Nature}, 451(7174):42, 2008.

\bibitem{Huse2003coulomb}
David~A. Huse, Werner Krauth, R.~Moessner, and S.~L. Sondhi.
\newblock Coulomb and liquid dimer models in three dimensions.
\newblock {\em Phys. Rev. Lett.}, 91:167004, Oct 2003.

\bibitem{Castelnovo2011debye}
C.~Castelnovo, R.~Moessner, and S.~L. Sondhi.
\newblock Debye-h\"uckel theory for spin ice at low temperature.
\newblock {\em Phys. Rev. B}, 84:144435, Oct 2011.

\bibitem{Kaiser2018emergent}
Vojt{\v{e}}ch Kaiser, Jonathan Bloxsom, Laura Bovo, Steven~T Bramwell, Peter~CW
  Holdsworth, and Roderich Moessner.
\newblock Emergent electrochemistry in spin ice: Debye-h$\backslash$"$\{$u$\}$
  ckel theory and beyond.
\newblock {\em arXiv preprint arXiv:1803.04668}, 2018.

\bibitem{Senthil2004deconfined}
T~Senthil, Ashvin Vishwanath, Leon Balents, Subir Sachdev, and Matthew~PA
  Fisher.
\newblock Deconfined quantum critical points.
\newblock {\em Science}, 303(5663):1490--1494, 2004.

\bibitem{Sandvik2007evidence}
Anders~W. Sandvik.
\newblock Evidence for deconfined quantum criticality in a two-dimensional
  heisenberg model with four-spin interactions.
\newblock {\em Phys. Rev. Lett.}, 98:227202, Jun 2007.

\bibitem{Broholm1990antiferromagnetic}
C.~Broholm, G.~Aeppli, G.~P. Espinosa, and A.~S. Cooper.
\newblock Antiferromagnetic fluctuations and short-range order in a kagom\'e
  lattice.
\newblock {\em Phys. Rev. Lett.}, 65:3173--3176, Dec 1990.

\bibitem{Moessner2011quantum}
Roderich Moessner and Kumar~S Raman.
\newblock Quantum dimer models.
\newblock In {\em Introduction to Frustrated Magnetism}, pages 437--479.
  Springer, 2011.

\bibitem{Knolle2014dynamics}
J.~Knolle, D.~L. Kovrizhin, J.~T. Chalker, and R.~Moessner.
\newblock Dynamics of a two-dimensional quantum spin liquid: Signatures of
  emergent majorana fermions and fluxes.
\newblock {\em Phys. Rev. Lett.}, 112:207203, May 2014.

\bibitem{Knolle2016dynamics}
Johannes Knolle.
\newblock {\em Dynamics of a Quantum Spin Liquid (Springer Theses)}.
\newblock Springer International Publishing, 1 edition, 2016.

\bibitem{Knolle2015dynamics}
J.~Knolle, D.~L. Kovrizhin, J.~T. Chalker, and R.~Moessner.
\newblock Dynamics of fractionalization in quantum spin liquids.
\newblock {\em Phys. Rev. B}, 92:115127, Sep 2015.

\bibitem{Lake2013multi}
B.~Lake, D.~A. Tennant, J.-S. Caux, T.~Barthel, U.~Schollw\"ock, S.~E. Nagler,
  and C.~D. Frost.
\newblock Multispinon continua at zero and finite temperature in a near-ideal
  heisenberg chain.
\newblock {\em Phys. Rev. Lett.}, 111:137205, Sep 2013.

\bibitem{Mourigal2013fractional}
Martin Mourigal, Mechthild Enderle, Axel Kl�pperpieper, Jean-S�bastien
  Caux, Anne Stunault, and Henrik~M. R�nnow.
\newblock Fractional spinon excitations in the quantum heisenberg
  antiferromagnetic chain.
\newblock {\em Nature Physics}, 9:435--441, 2013.

\bibitem{Banerjee2017neutron}
Arnab Banerjee, Jiaqiang Yan, Johannes Knolle, Craig~A Bridges, Matthew~B
  Stone, Mark~D Lumsden, David~G Mandrus, David~A Tennant, Roderich Moessner,
  and Stephen~E Nagler.
\newblock Neutron scattering in the proximate quantum spin liquid
  $\alpha$-rucl3.
\newblock {\em Science}, 356(6342):1055--1059, 2017.

\bibitem{Sandilands2015scattering}
Luke~J. Sandilands, Yao Tian, Kemp~W. Plumb, Young-June Kim, and Kenneth~S.
  Burch.
\newblock Scattering continuum and possible fractionalized excitations in
  $\ensuremath{\alpha}\text{\ensuremath{-}}{\mathrm{rucl}}_{3}$.
\newblock {\em Phys. Rev. Lett.}, 114:147201, Apr 2015.

\bibitem{Bulaevskii2008electronic}
L.~N. Bulaevskii, C.~D. Batista, M.~V. Mostovoy, and D.~I. Khomskii.
\newblock Electronic orbital currents and polarization in mott insulators.
\newblock {\em Phys. Rev. B}, 78:024402, Jul 2008.

\bibitem{Ng2007power}
Tai-Kai Ng and Patrick~A. Lee.
\newblock Power-law conductivity inside the mott gap: Application to
  $\ensuremath{\kappa}\mathrm{\text{\ensuremath{-}}}(\mathrm{BEDT}\mathrm{\text{\ensuremath{-}}}\mathrm{TTF}{)}_{2}{\mathrm{cu}}_{2}(\mathrm{CN}{)}_{3}$.
\newblock {\em Phys. Rev. Lett.}, 99:156402, Oct 2007.

\bibitem{Potter2013mechanism}
Andrew~C. Potter, T.~Senthil, and Patrick~A. Lee.
\newblock Mechanisms for sub-gap optical conductivity in herbertsmithite.
\newblock {\em Phys. Rev. B}, 87:245106, Jun 2013.

\bibitem{Huh2013optical}
Yejin Huh, Matthias Punk, and Subir Sachdev.
\newblock Optical conductivity of visons in ${Z}_{2}$ spin liquids close to a
  valence bond solid transition on the kagome lattice.
\newblock {\em Phys. Rev. B}, 87:235108, Jun 2013.

\bibitem{Bolens2017mechanism}
Adrien Bolens, Hosho Katsura, Masao Ogata, and Seiji Miyashita.
\newblock Mechanism for sub-gap optical conductivity in honeycomb kitaev
  materials.
\newblock {\em arXiv preprint arXiv:1711.00308}, 2017.

\bibitem{Little2017anti}
A.~Little, Liang Wu, P.~Lampen-Kelley, A.~Banerjee, S.~Patankar, D.~Rees, C.~A.
  Bridges, J.-Q. Yan, D.~Mandrus, S.~E. Nagler, and J.~Orenstein.
\newblock Antiferromagnetic resonance and terahertz continuum in
  $\ensuremath{\alpha}\text{\ensuremath{-}}{\mathrm{rucl}}_{3}$.
\newblock {\em Phys. Rev. Lett.}, 119:227201, Nov 2017.

\bibitem{Wang2017magnetic}
Zhe Wang, S.~Reschke, D.~H\"uvonen, S.-H. Do, K.-Y. Choi, M.~Gensch, U.~Nagel,
  T.~R\~o\ om, and A.~Loidl.
\newblock Magnetic excitations and continuum of a possibly field-induced
  quantum spin liquid in
  $\ensuremath{\alpha}\text{\ensuremath{-}}{\mathrm{rucl}}_{3}$.
\newblock {\em Phys. Rev. Lett.}, 119:227202, Nov 2017.

\bibitem{Wellm2017signatures}
C~Wellm, J~Zeisner, A~Alfonsov, AUB Wolter, M~Roslova, A~Isaeva, T~Doert,
  M~Vojta, B~B{\"u}chner, and V~Kataev.
\newblock Signatures of low-energy fractionalized excitations in $\alpha$-rucl
  $_3 $ from field-dependent microwave absorption.
\newblock {\em arXiv preprint arXiv:1710.00670}, 2017.

\bibitem{Devereaux2007inelastic}
Thomas~P. Devereaux and Rudi Hackl.
\newblock Inelastic light scattering from correlated electrons.
\newblock {\em Rev. Mod. Phys.}, 79:175--233, Jan 2007.

\bibitem{Cepas2008detection}
O.~C\'epas, J.~O. Haerter, and C.~Lhuillier.
\newblock Detection of weak emergent broken-symmetries of the kagome
  antiferromagnet by raman spectroscopy.
\newblock {\em Phys. Rev. B}, 77:172406, May 2008.

\bibitem{Ko2010raman}
Wing-Ho Ko, Zheng-Xin Liu, Tai-Kai Ng, and Patrick~A. Lee.
\newblock Raman signature of the u(1) dirac spin-liquid state in the
  spin-$\frac{1}{2}$ kagome system.
\newblock {\em Phys. Rev. B}, 81:024414, Jan 2010.

\bibitem{Knolle2014raman}
J.~Knolle, Gia-Wei Chern, D.~L. Kovrizhin, R.~Moessner, and N.~B. Perkins.
\newblock Raman scattering signatures of kitaev spin liquids in
  ${A}_{2}{\mathrm{iro}}_{3}$ iridates with $a=\mathrm{Na}$ or li.
\newblock {\em Phys. Rev. Lett.}, 113:187201, Oct 2014.

\bibitem{Perreault2015theory}
Brent Perreault, Johannes Knolle, Natalia~B. Perkins, and F.~J. Burnell.
\newblock Theory of raman response in three-dimensional kitaev spin liquids:
  Application to $\ensuremath{\beta}$- and
  $\ensuremath{\gamma}\ensuremath{-}{\mathrm{li}}_{2}{\mathrm{iro}}_{3}$
  compounds.
\newblock {\em Phys. Rev. B}, 92:094439, Sep 2015.

\bibitem{Perreault2016resonant}
Brent Perreault, Johannes Knolle, Natalia~B. Perkins, and F.~J. Burnell.
\newblock Resonant raman scattering theory for kitaev models and their majorana
  fermion boundary modes.
\newblock {\em Phys. Rev. B}, 94:104427, Sep 2016.

\bibitem{Maczka2008temperature}
M.~Maczka, M.~L. Sanju\'an, A.~F. Fuentes, K.~Hermanowicz, and J.~Hanuza.
\newblock Temperature-dependent raman study of the spin-liquid pyrochlore
  ${\text{tb}}_{2}{\text{ti}}_{2}{\text{o}}_{7}$.
\newblock {\em Phys. Rev. B}, 78:134420, Oct 2008.

\bibitem{Wulferding2010interplay}
Dirk Wulferding, Peter Lemmens, Patric Scheib, Jens R\"oder, Philippe Mendels,
  Shaoyan Chu, Tianheng Han, and Young~S. Lee.
\newblock Interplay of thermal and quantum spin fluctuations in the kagome
  lattice compound herbertsmithite.
\newblock {\em Phys. Rev. B}, 82:144412, Oct 2010.

\bibitem{Glamazda2016raman}
A.~Glamazda, P.~Lemmens, S.~H. Do, Y.~S. Choi, and K.~Y. Choi.
\newblock Raman spectroscopic signature of fractionalized excitations in the
  harmonic-honeycomb iridates $\beta$- and $\gamma$-li$_2$iro$_3$.
\newblock {\em Nature Communications}, 7:12286, 2016.

\bibitem{Halasz2016resonant}
G\'abor~B. Hal\'asz, Natalia~B. Perkins, and Jeroen van~den Brink.
\newblock Resonant inelastic x-ray scattering response of the kitaev honeycomb
  model.
\newblock {\em Phys. Rev. Lett.}, 117:127203, Sep 2016.

\bibitem{Natori2017dynamics}
W.~M.~H. Natori, M.~Daghofer, and R.~G. Pereira.
\newblock Dynamics of a $j=\frac{3}{2}$ quantum spin liquid.
\newblock {\em Phys. Rev. B}, 96:125109, Sep 2017.

\bibitem{Carretta2011nmr}
Pietro Carretta and Amit Keren.
\newblock Nmr and $\mu$sr in highly frustrated magnets.
\newblock In {\em Introduction to Frustrated Magnetism}, pages 79--105.
  Springer, 2011.

\bibitem{Yaouanc2011muon}
Alain Yaouanc and Pierre~Dalmas De~Reotier.
\newblock {\em Muon spin rotation, relaxation, and resonance: applications to
  condensed matter}, volume 147.
\newblock Oxford University Press, 2011.

\bibitem{Hsieh2008topological}
David Hsieh, Dong Qian, Lewis Wray, YuQi Xia, Yew~San Hor, Robert~Joseph Cava,
  and M~Zahid Hasan.
\newblock A topological dirac insulator in a quantum spin hall phase.
\newblock {\em Nature}, 452(7190):970, 2008.

\bibitem{Mourik2012Mourik}
V.~Mourik, K.~Zuo, S.~M. Frolov, S.~R. Plissard, E.~P. A.~M. Bakkers, and L.~P.
  Kouwenhoven.
\newblock Signatures of majorana fermions in hybrid
  superconductor-semiconductor nanowire devices.
\newblock {\em Science}, 336(6084):1003--1007, 2012.

\bibitem{Vasyukov2013scanning}
Denis Vasyukov, Yonathan Anahory, Lior Embon, Dorri Halbertal, Jo~Cuppens, Lior
  Neeman, Amit Finkler, Yehonathan Segev, Yuri Myasoedov, Michael~L Rappaport,
  et~al.
\newblock A scanning superconducting quantum interference device with single
  electron spin sensitivity.
\newblock {\em Nature nanotechnology}, 8(9):639, 2013.

\bibitem{Perreault2016majorana}
Brent Perreault, Stephan Rachel, F.~J. Burnell, and Johannes Knolle.
\newblock Majorana landau level raman spectroscopy.
\newblock {\em arXiv:1612.03951}, 2016.

\bibitem{Fransson2010theory}
J.~Fransson, O.~Eriksson, and A.~V. Balatsky.
\newblock Theory of spin-polarized scanning tunneling microscopy applied to
  local spins.
\newblock {\em Phys. Rev. B}, 81:115454, Mar 2010.

\bibitem{Coldea2010quantum}
R.~Coldea, D.~A. Tennant, E.~M. Wheeler, E.~Wawrzynska, D.~Prabhakaran,
  M.~Telling, K.~Habicht, P.~Smeibidl, and K.~Kiefer.
\newblock Quantum criticality in an ising chain: Experimental evidence for
  emergent e8 symmetry.
\newblock {\em Science}, 327(5962):177--180, 2010.

\bibitem{Wan2012quantum}
Yuan Wan and Oleg Tchernyshyov.
\newblock Quantum strings in quantum spin ice.
\newblock {\em Phys. Rev. Lett.}, 108:247210, Jun 2012.

\bibitem{Ivanov2001non}
D.~A. Ivanov.
\newblock Non-abelian statistics of half-quantum vortices in $\mathit{p}$-wave
  superconductors.
\newblock {\em Phys. Rev. Lett.}, 86:268--271, Jan 2001.

\bibitem{Smith2015neutron}
A.~Smith, J.~Knolle, D.~L. Kovrizhin, J.~T. Chalker, and R.~Moessner.
\newblock Neutron scattering signatures of the 3d hyperhoneycomb kitaev quantum
  spin liquid.
\newblock {\em Phys. Rev. B}, 92:180408, Nov 2015.

\bibitem{Smith2016majorana}
A.~Smith, J.~Knolle, D.~L. Kovrizhin, J.~T. Chalker, and R.~Moessner.
\newblock Majorana spectroscopy of three-dimensional kitaev spin liquids.
\newblock {\em Phys. Rev. B}, 93:235146, Jun 2016.

\bibitem{Morampudi2017statistics}
Siddhardh~C. Morampudi, Ari~M. Turner, Frank Pollmann, and Frank Wilczek.
\newblock Statistics of fractionalized excitations through threshold
  spectroscopy.
\newblock {\em Phys. Rev. Lett.}, 118:227201, May 2017.

\bibitem{Savary2017disorder}
Lucile Savary and Leon Balents.
\newblock Disorder-induced quantum spin liquid in spin ice pyrochlores.
\newblock {\em Phys. Rev. Lett.}, 118:087203, Feb 2017.

\bibitem{Sen2015topological}
Arnab Sen and R.~Moessner.
\newblock Topological spin glass in diluted spin ice.
\newblock {\em Phys. Rev. Lett.}, 114:247207, Jun 2015.

\bibitem{Hagiwara1990observation}
M.~Hagiwara, K.~Katsumata, Ian Affleck, B.~I. Halperin, and J.~P. Renard.
\newblock Observation of s=1/2 degrees of freedom in an s=1 linear-chain
  heisenberg antiferromagnet.
\newblock {\em Phys. Rev. Lett.}, 65:3181--3184, Dec 1990.

\bibitem{Mross2011charge}
David~F. Mross and T.~Senthil.
\newblock Charge friedel oscillations in a mott insulator.
\newblock {\em Phys. Rev. B}, 84:041102, Jul 2011.

\bibitem{Bhatt1982scaling}
R.~N. Bhatt and P.~A. Lee.
\newblock Scaling studies of highly disordered spin-1/2 antiferromagnetic
  systems.
\newblock {\em Phys. Rev. Lett.}, 48:344--347, Feb 1982.

\bibitem{Damle2000dynamics}
Kedar Damle, Olexei Motrunich, and David~A. Huse.
\newblock Dynamics and transport in random antiferromagnetic spin chains.
\newblock {\em Phys. Rev. Lett.}, 84:3434--3437, Apr 2000.

\bibitem{Willans2011site}
A.~J. Willans, J.~T. Chalker, and R.~Moessner.
\newblock Site dilution in the kitaev honeycomb model.
\newblock {\em Phys. Rev. B}, 84:115146, Sep 2011.

\bibitem{Kimchi2017valence}
Itamar Kimchi, Adam Nahum, and T~Senthil.
\newblock Valence bonds in random quantum magnets: Theory and application to
  ybmggao4.
\newblock {\em arXiv preprint arXiv:1710.06860}, 2017.

\bibitem{Hentrich2018unusual}
Richard Hentrich, Anja U.~B. Wolter, Xenophon Zotos, Wolfram Brenig, Domenic
  Nowak, Anna Isaeva, Thomas Doert, Arnab Banerjee, Paula Lampen-Kelley,
  David~G. Mandrus, Stephen~E. Nagler, Jennifer Sears, Young-June Kim, Bernd
  B\"uchner, and Christian Hess.
\newblock Unusual phonon heat transport in
  $\ensuremath{\alpha}\text{\ensuremath{-}}{\mathrm{rucl}}_{3}$: Strong
  spin-phonon scattering and field-induced spin gap.
\newblock {\em Phys. Rev. Lett.}, 120:117204, Mar 2018.

\bibitem{Banerjee2018excitations}
Arnab Banerjee, Paula Lampen-Kelley, Johannes Knolle, Christian Balz,
  Adam~Anthony Aczel, Barry Winn, Yaohua Liu, Daniel Pajerowski, Jiaqiang Yan,
  Craig~A Bridges, et~al.
\newblock Excitations in the field-induced quantum spin liquid state of
  $\alpha$-rucl 3.
\newblock {\em npj Quantum Materials}, 3(1):8, 2018.

\bibitem{Dean2016ultrafast}
MPM Dean, Yue Cao, X~Liu, S~Wall, D~Zhu, Roman Mankowsky, V~Thampy, XM~Chen,
  JG~Vale, D~Casa, et~al.
\newblock Ultrafast energy-and momentum-resolved dynamics of magnetic
  correlations in the photo-doped mott insulator sr 2 iro 4.
\newblock {\em Nature materials}, 15(6):601, 2016.

\bibitem{Alpichshev2015confinement}
Zhanybek Alpichshev, Fahad Mahmood, Gang Cao, and Nuh Gedik.
\newblock Confinement-deconfinement transition as an indication of
  spin-liquid-type behavior in na 2 iro 3.
\newblock {\em Physical review letters}, 114(1):017203, 2015.

\bibitem{Yamada2017designing}
Masahiko~G. Yamada, Hiroyuki Fujita, and Masaki Oshikawa.
\newblock Designing kitaev spin liquids in metal-organic frameworks.
\newblock {\em Phys. Rev. Lett.}, 119:057202, Aug 2017.

\bibitem{Nakatsuji2006metallic}
S.~Nakatsuji, Y.~Machida, Y.~Maeno, T.~Tayama, T.~Sakakibara, J.~van Duijn,
  L.~Balicas, J.~N. Millican, R.~T. Macaluso, and Julia~Y. Chan.
\newblock Metallic spin-liquid behavior of the geometrically frustrated kondo
  lattice ${\mathrm{pr}}_{2}{\mathrm{ir}}_{2}{\mathrm{o}}_{7}$.
\newblock {\em Phys. Rev. Lett.}, 96:087204, Mar 2006.

\bibitem{Duan2003controlling}
L.-M. Duan, E.~Demler, and M.~D. Lukin.
\newblock Controlling spin exchange interactions of ultracold atoms in optical
  lattices.
\newblock {\em Phys. Rev. Lett.}, 91:090402, Aug 2003.

\bibitem{Manmana2013topological}
Salvatore~R. Manmana, E.~M. Stoudenmire, Kaden R.~A. Hazzard, Ana~Maria Rey,
  and Alexey~V. Gorshkov.
\newblock Topological phases in ultracold polar-molecule quantum magnets.
\newblock {\em Phys. Rev. B}, 87:081106, Feb 2013.

\end{thebibliography}

\end{document}